\documentclass[sigconf]{acmart}
\renewcommand\footnotetextcopyrightpermission[1]{} 
\usepackage{graphicx}
\usepackage{textcomp}
\usepackage{xcolor}

\usepackage{booktabs} 
\usepackage{float}
\usepackage{multirow}
\usepackage{indentfirst}
\usepackage{multirow}
\usepackage{amsmath}
\usepackage{graphicx}
\usepackage{subfigure}
\usepackage[]{graphicx,indentfirst}
\usepackage{arydshln}
\usepackage{textcomp}
\usepackage{multicol}
\usepackage{subfigure}
\usepackage{bm}
\usepackage{url}
\usepackage{amsfonts}
\usepackage{mathrsfs}
\usepackage{algorithmic}
\usepackage[ruled,linesnumbered]{algorithm2e}

\usepackage{ulem}

\begin{document}

\title{Self-supervised Contrastive Learning for Implicit Collaborative Filtering}

\author{
	{Shipeng Song$^{1}$, Bin Liu$^{*,2}$, Fei Teng$^{2}$, Tianrui Li$^{2}$}\\
	{
    {$^{1}$School of Electronic Information and Communications, Huazhong University of
    	Science and Technology (HUST), Wuhan \\
    	$^{2}$School of Computing and Artificial Intelligence, Southwest Jiaotong University, Chendu}}\\
}
	
\begin{abstract}
Contrastive learning-based recommendation algorithms have significantly advanced the field of self-supervised recommendation, particularly with BPR as a representative ranking prediction task that dominates implicit collaborative filtering. However, the presence of false-positive and false-negative examples in recommendation systems hampers accurate preference learning. In this study, we propose a simple self-supervised contrastive learning framework that leverages positive feature augmentation and negative label augmentation to improve the self-supervisory signal. Theoretical analysis demonstrates that our learning method is equivalent to maximizing the likelihood estimation with latent variables representing user interest centers. Additionally, we establish an efficient negative label augmentation technique that samples unlabeled examples with a probability linearly dependent on their relative ranking positions, enabling efficient augmentation in constant time complexity. Through validation on multiple datasets, we illustrate the significant improvements our method achieves over the widely used BPR optimization objective while maintaining comparable runtime. 

\end{abstract}


\maketitle

\section{Introduction} \label{Introduction}
In the rapidly evolving information age of the Internet, recommendation systems play a crucial role in various domains such as e-commerce, social media, and news reading. However, obtaining accurate user preference rating labels for supervised model training is both expensive and challenging, given the inherent characteristics of recommendation system scenarios. Currently, many state-of-the-art recommendation models leverage self-supervised learning (SSL) techniques~\cite{Liu:2021:TKDE,Chen:2020:NIPS,BYOL:2020:NIPS,wu:2023:TKDE,SSR:2023:TKDE} to design self-supervised tasks for (pre-)training the models to encode user-item feature representations, which are then used for predicting personalized ranking (downstream tasks). For instance, generative self-supervised tasks train recommendation models to reconstruct input data~\cite{sun2019bert4rec,geng2022recommendation,zhang:sigir}, while contrastive self-supervised tasks train models to encode differences between positive and negative samples~\cite{Steffen:2009:UAI,Xiangnan:2020:SIGIR,Wang:2019:SIGIR,Jingtao:2019:IJCAI,lightgcl:2023:ICLR}. These recommendation models trained with self-supervised tasks are also referred to as self-supervised recommendation (SSR) models, representing an important application area of self-supervised learning. SSR fully exploits the inherent relationships in the input data to extract self-supervised signals, which not only reduces the reliance on manually annotated datasets but also helps alleviate issues in traditional recommendation systems such as data sparsity, fake correlations, and adversarial attacks~\cite{Liu:2021:TKDE,SSR:2023:TKDE}.

Recommendation algorithms based on contrastive learning (CL)~\cite{Oord:2018:arxiv,Gutmann:2010:ICAIS,Wang:2020:ICML}are the primary form of implementation for self-supervised recommendation~\cite{SSR:2023:TKDE}, significantly advancing the forefront of self-supervised recommendation and driving extensive attention~\cite{liu2021contrastive,10.1145/3477495.3532009,qiu2022contrastive,yu2023xsimgcl}. Among them, the pioneering approach is Bayesian Personalized Ranking (BPR)~\cite{Steffen:2009:UAI}, which draws inspiration from self-supervised learning techniques and automatically labels items as positive or negative based on the presence or absence of interactions. The designed BPR loss is, in fact, a special case of contrastive loss with a negative sample count of 1 and is also referred to as pairwise loss. On the one hand, by optimizing the contrastive loss, the distance between positive examples and anchor points is minimized while maximizing the distance between negative examples and anchor points, thereby optimizing mutual information~\cite{Oord:2018:arxiv}. On the other hand, optimizing the contrastive loss leads to improvements in ranking metrics such as Area Under the Curve (AUC) and Normalized Discounted Cumulative Gain (NDCG)~\cite{Steffen:2009:UAI,Jiancan:2022:arxiv}. As BPR is optimized specifically for ranking objectives, it gradually dominates the field of implicit collaborative filtering.


Although these methods have demonstrated remarkable generalization capabilities within self-supervised setting, there is a significant gap compared to contrastive recommendation algorithms in a supervised setting~\cite{SSR:2023:TKDE,Chuang:2020:NIPS}: the positive and negative sample labels used for preference comparison are "pseudo-labels" rather than true preference labels of "liking" or "disliking". In the context of recommendation systems, the semantics of positive examples are "liked items," while negative examples represent "disliked items." However, users typically express their preferences or interests only through interaction behaviors such as clicks or purchases. On the one hand, these interaction behaviors often lack explicit ratings. Interactions that do not reflect preferences, such as proxy purchases, accidental clicks, or mere views, cannot be distinguished from normal interactions and are all labeled as positive examples~\cite{Wang:2021:WSDM}. On the other hand, as users only provide positive feedback, it is impossible to observe which items they dislike. Thus, all items that users have not seen are labeled as negative examples, but they are actually unlabeled data~\cite{Zhang:2013:SIGIR,Yang:2020:KDD}. Therefore, the datasets used for training recommendation systems often exist in a positive-unlabeled (PU) form, where positive examples are also contaminated with noise. Self-supervised contrastive learning methods, represented by BPR, automatically label interaction data as positive or negative based on the presence or absence of observations, which leads to the issue of pseudo-positive and pseudo-negative examples, adversely affecting the learning of accurate user-item feature representations.


To address the issue of pseudo-positive examples, previous research has primarily focused on optimizing strategies, with an emphasis on exploring the dynamics of the optimization process to tackle label noise learning problems. The core insight is that neural networks tend to first fit clean data and then fit noisy data~\cite{zhang2021understanding}. Building upon this finding, DenoiseRec~\cite{Wang:2021:WSDM} pioneered the introduction of the small loss trick in recommendation systems. The core idea is that samples with higher loss values are more likely to be noise, and a dynamic threshold function is used to truncate the loss values. Empirical observations based on optimizing strategies are primarily focused on losses constructed from individual samples, and it remains uncertain whether they still hold for losses constructed from paired samples, as difficult-to-fit pairs often contribute more to performance improvement in recommendations~\cite{Steffen:2014:WSDM}.

Regarding the issue of pseudo-negative examples, previous research has primarily focused on designing negative sampling algorithms based on side information such as social relationships and gender, as well as model information such as scores and ranking positions, to sample negative examples from unlabeled data. Existing negative sampling algorithms can be classified into static negative sampling~\cite{Steffen:2009:UAI} and dynamic negative sampling~\cite{Steffen:2014:WSDM,Ding:2020:NIPS,Zhang:2013:SIGIR,Jingtao:2019:IJCAI,Yang:2020:KDD,Bin:2023:ICDE}, depending on whether the sampling distribution changes with the model's state. Static negative sampling algorithms adopt fixed sampling distributions, such as uniform sampling or sampling based on item popularity. On the other hand, dynamic negative sampling algorithms adjust the sampling distribution continuously during model training to sample negative examples that are more similar to positive examples in the embedding space, i.e., hard negative samples that are difficult to distinguish. For example, samples with higher scores or higher rankings can be sampled. However, static negative sampling algorithms heavily rely on side information as supervision signals, while dynamic negative sampling introduces higher computational costs and is prone to sampling pseudo-negative examples~\cite{Bin:2023:ICDE}.
\begin{figure}[!h]
	\centering
	\includegraphics[width=0.35\textwidth]{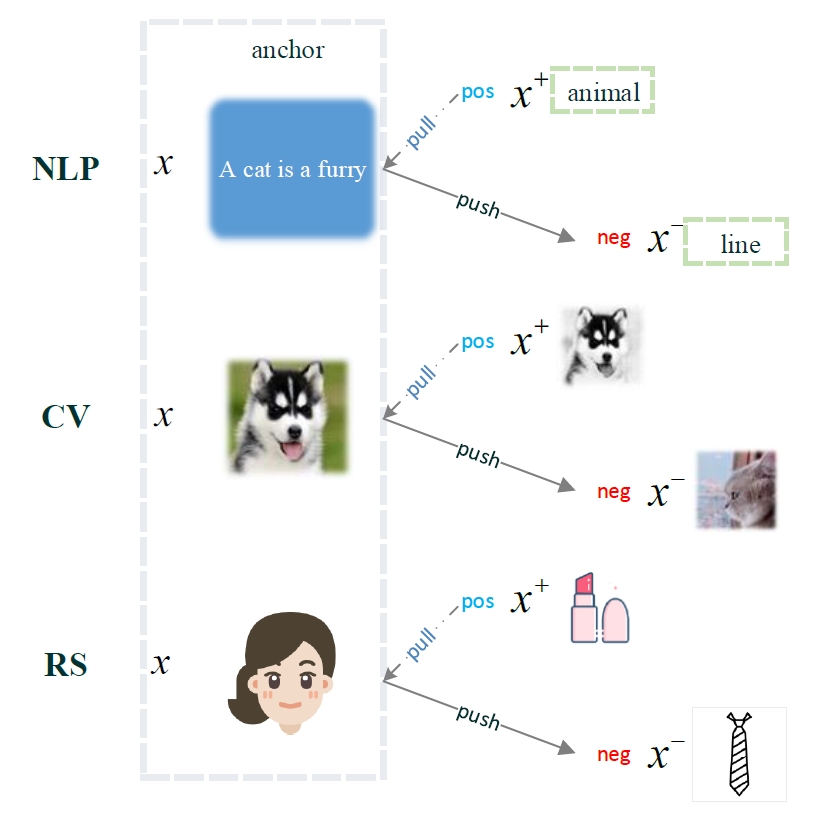}
	\caption{Various applications of contrastive learning.}
	\label{2Fig:illustrative}
\end{figure}

Building upon the work of BPR, this study proposes a simple and easily implementable self-supervised contrastive learning framework for implicit collaborative filtering. By considering both positive instance feature augmentation and negative instance label enrichment, it comprehensively addresses the issues of pseudo-positive and pseudo-negative examples, aiming to obtain more informative self-supervised signals and learn more accurate user-item feature representations. The proposed method is straightforward and practical, achieving significant improvements in top-k ranking accuracy on four publicly available datasets while maintaining a strict linear time complexity relative to BPR.

\section{Related Work}
\subsection{Self-supervised Recommendation}

Self-supervised learning (SSL)~\cite{Liu:2021:TKDE,SSR:2023:TKDE} is a powerful approach that leverages carefully designed pretext tasks to extract transferable knowledge from unlabeled data. By utilizing transformations or processing of the input data, SSL generates supervised signals, reducing the need for manual labels. This methodology has gained significant popularity in recommendation systems. A prominent example is Bayesian Personalized Ranking (BPR)~\cite{Steffen:2009:UAI}, where interacted items are automatically labeled as positive examples, while negative examples represent non-interacted items. BPR employs a self-supervised signal that encourages the model to assign higher scores to positive examples compared to negative examples during the optimization process. 

Self-supervised recommendation can mainly be categorized into contrastive recommendation algorithms, generative recommendation algorithms, and hybrid recommendation algorithms. Contrastive recommendation algorithms~\cite{Xiangnan:2020:SIGIR,10.1145/3404835.3462862,yu2023xsimgcl}, represented by BPR, are the primary form of self-supervised recommendation. Their core idea is to contrast  positive and negative samples, aiming for the model to assign higher scores to positive examples than negative examples, thus optimizing the AUC of ranking list. Recently, graph-based contrastive recommendation algorithms have also emerged~\cite{yu2023xsimgcl,lightgcl:2023:ICLR}, where the self-supervised signal is derived from the semantic invariance resulting from slight perturbations of the user-item bipartite graph. However, such tasks are inconsistent with ranking tasks and often require joint optimization with the BPR loss~\cite{lightgcl:2023:ICLR,yu2023xsimgcl}. Generative recommendation algorithms~\cite{sun2019bert4rec,geng2022recommendation,zhang:sigir} aim to reconstruct the original input by using perturbed input, thereby encoding the intrinsic correlations within the data. Hybrid methods~\cite{bian2021contrastive,wang2023curriculum} leverage different self-supervised signals and combine multiple types of pretraining tasks to obtain enhanced and comprehensive self-supervised signals, which often outperform single self-supervised signals in training effectiveness. However, hybrid methods face the challenge of coordinating multiple self-supervised tasks and usually require manual hyperparameter search to balance the different self-supervised tasks.

\subsection{Contrastive Learning}
The paradigm of learn-to-compare~\cite{Gutmann:2010:ICAIS} has achieved significant success in various fields such as computer vision, natural language processing, and recommendation systems, and has proven beneficial for almost all types of downstream tasks~\cite{Bachman:2019:NIPS,He:2020:CVPR,Huang:2019:ICML,Wu:2018:CVPR,luo2023segclip,Chu:2023:PAMI}. While the encoders and similarity metrics may vary depending on the task~\cite{Dosovitskiy:2014:NIPS,Wenqi:2021:KDD,10.1145/3543507.3583251}, contrastive learning shares a common fundamental idea of optimizing contrastive losses~\cite{Oord:2018:arxiv,Gutmann:2010:ICAIS,Wang:2020:ICML,Hjelm:2018:Arxiv}, which involve pulling positive samples closer and pushing negative samples farther apart, indirectly optimizing the lower bound of mutual information between variables ~\cite{Oord:2018:arxiv}. Particularly in the domain of personalized recommendation, optimizing contrastive losses leads to the optimization of ranking metrics such as AUC and NDCG for the ranking list~\cite{Steffen:2009:UAI,Jiancan:2022:arxiv}. This desirable property has made contrastive learning widely applicable in ranking prediction tasks.

Contrastive learning is a discriminative approach that aims to learn through comparisons by utilizing the objective function of Noise Contrastive Estimation (NCE) ~\cite{Gutmann:2010:ICAIS}
\begin{eqnarray}
	\mathcal{L}_\textsc{Nce} = \mathbb E[-\log \frac{\exp(f(x)^Tf(x^+))}{\exp(f(x)^Tf(x^+))+\exp(f(x)^Tf(x^-))}]
\end{eqnarray}

Among these, $x$ serves as the anchor point, typically representing a user $u$ in the context of recommendation. $x^+$ denotes the positive example for the anchor, often chosen as an interacted item $i$. Conversely, $x^-$ represents the negative example for the anchor, typically chosen as a non-interacted item $j$. The encoder $f$ is responsible for encoding the anchor point, and $f(x)^Tf(x^+)$ represents the predicted similarity score for the positive example, parameterized by the encoder. This score is often computed using inner product similarity or cosine similarity and denoted as $\hat{x}_{ui}$. Similarly, $f(x)^Tf(x^-)$ represents the predicted similarity score for the negative example, denoted as $\hat{x}_{uj}$. It can be observed that there is a connection between the Noise Contrastive Estimation (NCE) loss function and the Bayesian Personalized Ranking (BPR) loss~\cite{Steffen:2009:UAI}:

\begin{eqnarray}
	\mathcal{L}_\textsc{Nce} 
	&=& \mathbb E[-\log\frac{\exp(\hat{x}_{ui})}{\exp(\hat{x}_{ui})+\exp(\hat{x}_{uj})}]\nonumber \\
	&=& \mathbb E[-\log\frac{1}{1+\exp(\hat{x}_{uj}-\hat{x}_{ui})}]\nonumber \\
	&=& \mathbb E[-\log \sigma (\hat{x}_{ui} - \hat{x}_{uj})]\nonumber \\
	&=& \mathcal{L}_\textsc{Bpr}  \label{eq:bpr}
\end{eqnarray}

Here, $\sigma(\cdot)$ denotes the sigmoid function. In practice, regularization terms are introduced to prevent overfitting, and these regularization terms are equivalent to the logarithm of the prior density of a Gaussian distribution. Consequently, Bayesian Personalized Ranking (BPR) interprets the above equation as the maximum a posteriori estimation of the observed ordered pairs. Despite having different interpretations, BPR and Noise Contrastive Estimation (NCE) share identical mathematical forms and have the same optimization objective. By introducing additional negative examples into Noise Contrastive Estimation (NCE), one can obtain the InfoNCE loss~\cite{Oord:2018:arxiv}:
\begin{small}
	\begin{eqnarray} \label{eq:infonce}
		\mathcal{L}_\textsc{InfoNCE} = \mathbb E[-\log \frac{\exp(f(u)^Tf(i))}{\exp(f(u)^Tf(i))+\sum_{n=1}^{N}\exp(f(u)^Tf(j_n))}]
	\end{eqnarray}
\end{small}

It can be seen that both NCE and BPR are specific cases of the InfoNCE loss when the number of negative examples, denoted as $N$, is equal to 1. Taking the most general form of the InfoNCE loss as an example, it represents the cross-entropy of correctly classifying positive samples~\cite{Oord:2018:arxiv}. The minimum value of the InfoNCE loss is 0, which is achieved when the similarity score between the positive example and the negative example satisfies $f(u)^Tf(i) - f(u)^Tf(j_n) \rightarrow +\infty$, where $n \in {1,2,\cdots N}$. This conclusion also applies to NCE and BPR, with the only difference being that the number of negative examples, $N$, is equal to 1.

\section{Methodology}
\subsection{Prelimiaries}
We begin by summarizing the prevailing paradigm of ranking oriented collaborative filtering models based on contrastive learning. We consider $\mathcal{U}$ as the set of users and $\mathcal{I}$ as the set of items. $\mathcal{I}_u^+$ indicates the positive item set that user $u$ has previously interacted with, while $\mathcal{I}_u^-$ indicates the negative item set that user $u$ has not interacted with. Then $(u, i, j)$ constitutes a training triple, where $u \in \mathcal{U}, i\in \mathcal{I}_u^+, j\in \mathcal{I}_u^- $, and the corresponding training triplet set is denoted as $\mathcal{D}$.

Let $\hat{x}_{ui}$ represents the similarity score, parameterized by neural networks, indicating the predicted preference level of user $u$ for positive item $i$. Similarly, $\hat{x}_{uj}$ represents the similarity score indicating the predicted preference level of user $u$ for negative item $j$. Then the widely used Bayesian Personalized Ranking (BPR) loss of Eq.~\eqref{eq:bpr} is often implemented by the following empirical form in practice:
\begin{eqnarray}
	\mathcal{L}_\textsc{Bpr} = \frac{1}{|\mathcal{D}|} \sum_{(u,i,j) \in \mathcal{D}} -\log \sigma (\hat{x}_{ui} - \hat{x}_{uj})
\end{eqnarray}

\subsection{Positive Feature Augmentation}
The Bayesian Personalized Ranking (BPR) utilizes a self-supervised signal that states "the user's preference for positive examples is greater than negative examples," indicating that the model is optimized for assigning higher scores to positive examples than to negative ones. However, in the context of implicit collaborative filtering, positive examples (interacted items) do not necessarily represent positive preferences. Certain abnormal interaction patterns such as proxy purchases, accidental clicks, or mere views do not reflect user preferences but are recorded indiscriminately with other normal interactions, giving rise to the problem of false-positive examples. The issue  arises due to the incompleteness of implicit feedback data, where only the presence or absence of interactions can be observed, without capturing the magnitude of preference labels.

The erroneous self-supervised signal constructed based on false-positive examples can lead to overfitting to the data. The model may perceive false-positive examples as genuine user preference patterns, resulting in erroneous learning and predictions. More specifically, the model treats pseudo-positive examples as items that users like, causing the collaborative filtering mechanism to recommend similar items to false-positive examples and consequently leading to a high false-positive rate in the top-k recommendation list.

To mitigate the erroneous self-supervised signal caused by false-positive examples, we introduce the notion of interest center for user $u$ as a substitute for the original positive examples in the contrast with negative examples. Specifically, the interest center for user $u$ can be computed using $M$ positive samples as follows:
\begin{eqnarray}\label{eq:featureaug}
	\mathbf{p} = \frac{1}{M} \sum_{i \in \mathcal{I}_u^+} f(i),
\end{eqnarray}
where $f$ is a neural network that maps samples to a $d$-dimensional embedding: $f: i\in \mathcal{I} \rightarrow  f(i) \in \mathbb{R}^d$. Therefore, $f(i)$ represents the feature representation of the positive example in the embedding space. Then the meaning of vector $\mathbf{z}$ is the center of the positive representations in the embedding space, which corresponds to the \textit{interest center}. The similarity between a user and the interest center can be computed as follows:
\begin{eqnarray}\label{eq:possim}
	\hat{x}_{up} = \cos(f(u),\mathbf{p})/\tau,
\end{eqnarray}
where $\cos(\cdot,\cdot)$ measures the cosine similarity of user embedding and his \textit{interest center}, $\tau$ is temperature scaling. 

The core idea of our method is to replace the individual positive example $i$ with the  class center of $M$ positive examples when computing the similarity scores. This approach brings the user closer to their interest center in the contrastive learning process, rather than focusing on the distance to individual positive examples, thereby mitigating the adverse effects of false-positive examples on preference learning. Compared to individual positive examples, the interest center is smoother as it is composed of $M$ positive examples. Therefore, it is less influenced by false-positive examples and helps to learn more accurate user preferences. This idea aligns with previous findings in the field of contrastive learning, which suggest that the most representative positive sample contributes the most to the model's performance \cite{joshi2023data}. Intuitively, using the most representative positive sample helps the model learn the general patterns of user interests. The interest center constructed through Eq.~\eqref{eq:featureaug} is, in fact, the most representative positive sample. However, unlike prototype contrast, which involves clustering all items in the E-step, our method only enhances the features within the positive examples of the user to construct the interest center.

\subsection{Negative Label Augmentation}
In the case of negative examples (non-interacted items), these examples do not necessarily represent negative preferences. Most non-interacted items are items that users have not seen. Among these unobserved items, there are potential items of interest to the users, which are referred to as false-negative examples. In fact, the false-negative examples arises from the incompleteness of implicit feedback data. Automatic labeling of negative examples based on whether they are interacted with or not differs from the negative examples that users do not like. As a result, the model cannot accurately capture the preference  patterns of users, leading to unreliable prediction results. These pseudo-negative examples can cause the model to misjudge that users do not like such items, leading to a reduction in recommendations similar to the pseudo-negative examples by the collaborative filtering mechanism. Consequently, this leads to a lower true-positive rate in the top-k recommendation list.

Sampling a negative example from unlabeled samples involves label augmentation from the perspective of self-supervised learning, where an unlabeled sample is assigned a negative label for model training. The core idea behind label augmentation for unlabeled samples in this paper is that, difficult negative samples that are similar to positive examples are often already exposed to the user. In other words, the user has already observed these items but has not interacted with them, indicating a negative preference towards these items~\cite{Jingtao:2019:IJCAI}. This suggests that the probability of enhancing the label of an unlabeled sample as a negative example should be higher for samples that are more similar to positive examples.

This idea aligns with the core principle of existing mainstream methods for hard negative sampling. However, the motivation behind it differs. The motivation for hard negative sampling is primarily based on the notion that difficult negative samples contribute to larger loss values, resulting in larger gradient values. This allows the model to learn more information from such difficult negative samples, leading to greater gains in ranking performance. However, traditional hard negative sampling methods often involve significant computational costs. For instance, the negative sampling algorithm proposed by the authors of BPR assigns sampling probabilities proportional to the ranking positions of items, which introduces a substantial computational overhead due to the large number of items. Inspired by the accept-reject sampling algorithm, we randomly sample two samples $j$ and $j^\prime$ from unlabeled samples, and then with a probability of $\alpha$, we label the sample with the higher score as a negative example:
\begin{eqnarray}\label{eq:la1}
	\mathbb p \left(q = \arg \max_{q \in \{j,j^\prime\}} \hat{x}_{uq}\right) = \alpha
\end{eqnarray}
This means that with a probability of $1-\alpha$, we label the sample with the lower score as a negative example:
\begin{eqnarray}\label{eq:la2}
	\mathbb p \left(q = \arg \min_{q \in \{j,j^\prime\}} \hat{x}_{uq}\right) = 1-\alpha
\end{eqnarray}

\subsection{Optimization Criterion}
After the feature augmentation for positive examples and label augmentation for negative examples, we refine the self-supervised signal as: "the preference of the user towards the most representative positive example (interest center) being greater than the preference towards difficult negative samples (items that have been exposed to the user but are not liked by the user)". This is achieved by optimizing the following objective:
\begin{eqnarray}\label{eq:ssrobj}
	\mathcal{L} = \frac{1}{N} \sum-\log \sigma(\hat{x}_{up} - \hat{x}_{uq} )
\end{eqnarray}
where $N$ is the number of training triples. Thus, the optimization involves searching for an optimal set of parameters to maximize the following expression:
\begin{eqnarray}\label{eq:theta}
	\theta ^* = \arg \max \sum \log \sigma(\hat{x}_{up} - \hat{x}_{uq} )
\end{eqnarray}

The modified the self-supervised signal for ranking-oriented contrastive loss is easy to implement. The key lies in using representative positive examples as the interest center instead of the original single positive example, and using difficult negative examples obtained through label augmentation in place of random negative samples. The complete algorithm workflow is as depicted in Algorithm~\ref{alg:aunce}.

Complexity: Let us begin by examining the complexity of the baseline method, BPR, which is dependent on neural network $f(\cdot)$. For instance, considering matrix factorization with a latent dimension of $d$, we can easily extend this analysis to other models. Suppose we have a mini-batch data with a batch size of $bs$, composed of training triples $(u, i, j)$. During the forward propagation, a total of $2 \times bs$ item score predictions are made, resulting in a time complexity of $\mathcal O(2bs \times d)$. In the backward propagation, at most $3 \times bs$ embeddings are updated, involving a total of $5bs \times d$ operations, leading to a time complexity of $\mathcal O(bs \times d)$.

Compared to the standard BPR, the additional complexity of our proposed method lies in the computation of Eq.~\eqref{eq:featureaug} , which has a complexity of $\mathcal O (M\times d)$, where M is a hyperparameter, which is typically set to a small constant. As well as the computation of Eq.~\eqref{eq:la1} and Eq.~\eqref{eq:la2}, which have a complexity of O(1). Consequently, our method maintains a strict linear complexity relative to BPR.

\begin{algorithm}[!]
	\caption{The proposed learning algorithm}\label{alg:aunce}
	\KwIn{Interaction set, feature representation function $f(\cdot)$, number of $M$ for feature augmation, $\alpha$ for label augmation.}
	\KwOut{Parameters $\Theta$.}
	\For{\text{user} $u \in \mathcal{U}$}{
		~~ $\#$ \textit{feature augmation for positive sample} \\
		Uniformly sample  $M$ positive item $\{i_m\} _{m=1}^M \in \mathcal{I}_u^+$;\\
		Calculate representive positive item Via Eq.~\eqref{eq:featureaug}; \\
		Calculate positive simimilarity Via Eq.~\eqref{eq:possim}; \\
		~~  $\#$ \textit{ label augmation for negative sample} \\
		Uniformly sample  $2$ negative items $\{j, j^\prime \}\in \mathcal{I}_u^-$;\\
		prob = random.uniform(0,1); \\
		\If{prob $\leq \alpha$}{$q = \arg \max_{q \in \{j,j^\prime\}} \hat{x}_{uq}$} 
		\Else{$q = \arg \min_{q \in \{j,j^\prime\}} \hat{x}_{uq}$}
		Calculate loss via Eq.~\eqref{eq:ssrobj};\\
		Update $\Theta$.
	}
	\KwResult{Parameters $\Theta$.}
\end{algorithm}

\subsection{Theoretical Analysis}
In this paper, we obtain more dense and reliable self-supervised signals through feature augmentation for positive examples and label augmentation for negative examples. While being easy to implement, these techniques provide rich theoretical insights. In the following, we present two theoretical results to help readers better understand the positive feature augmentation and negative label augmentation employed in this study.

\begin{lemma}\label{lemma:1}
	Defining the likelihood of preferring positive examples over negative examples as a sigmoid function, denoted as $p(\succ;\theta) = \sigma(\cdot)$ in the context of BPR~\cite{Steffen:2009:UAI}. Then the solution obtained from Eq.~\eqref{eq:theta} is equivalent to maximizing the likelihood estimation for the latent variables $\mathbf p$ to be the centers of interacted item representations.
\end{lemma}
\textbf{Proof}: BPR (Bayesian Personalized Ranking) models the likelihood of a user preferring positive examples over negative examples using the sigmoid function:
\begin{eqnarray}
	p(\succ;\theta) = \sigma(\hat{x}^+ -\hat{x}^-)
\end{eqnarray}
where $\succ$ represents the observed positive and negative sample pairs, i.e., the preference structure, and $\theta$ denotes the model parameters of the neural network. We can rewrite Eq.~\eqref{eq:ssrobj} as follows:
\begin{eqnarray}
	\theta ^* &=& \arg \max \sum \log p(\succ;\theta)\\
	&=& \arg \max \sum \log \sum_{\mathbf{p}_i}  p(\succ,\mathbf{p}_i;\theta)
\end{eqnarray}
Optimizing this function directly is challenging, thus EM algorithm employ a surrogate function to provide a lower bound:
\begin{small}
	\begin{eqnarray}
		\sum \log\sum_{\mathbf{p}_i}  p(\succ,\mathbf{p}_i;\theta)  &=& \sum \log\sum_{\mathbf{p}_i} Q(\mathbf{p}_i)  \frac{p(\succ,\mathbf{p}_i;\theta)}{Q(\mathbf{p}_i)} \\
		&\geq &\sum \sum_{\mathbf{p}_i} Q(\mathbf{p}_i) \log \frac{p(\succ,\mathbf{p}_i;\theta)}{Q(\mathbf{p}_i)} 
	\end{eqnarray}
\end{small}

Therefore, maximizing the lower bound is equivalent to find $\theta^*$ that maximizes the following Q-function~\cite{Dempster:1977:RSS}:
\begin{eqnarray}
	\sum \sum_{\mathbf{p}_i} Q(\mathbf{p}_i)\log p(\succ,\mathbf{p}_i;\theta)
\end{eqnarray}
In the E-step, the model parameters $\theta$, i.e., user-item representations, are fixed, and we solve for $\mathbf{p}_i$. Since the user's interest representation  is defined as the center of the interacted items. Then, in the E-step, the center of $M$ positive embeddings is calculated as follows:
\begin{eqnarray}\label{eq:estep}
	\mathbf{p}_i = \frac{1}{M} \sum_{i \in \mathcal{I}_u^+} f(i),
\end{eqnarray}
In the M-step, with the latent variables $\mathbf{p}_i$ fixed, we solve for the optimal parameters that maximize the Q function
\begin{eqnarray}
	\theta ^* &=& \arg \max \sum \sum_{\mathbf{p}_i} Q(\mathbf{p}_i)\log p(\succ,\mathbf{p}_i;\theta) \\
	&=& \arg \max \sum \sum_{z_i} \log p(\succ,\mathbf{p}_i;\theta)  \label{eq:M2}\\
	&=& \arg \max \sum \log \sigma(\hat{x}_{up} - \hat{x}_{uq} ) \label{eq:M3}
\end{eqnarray}
Eq.~\eqref{eq:M2} is derived from the definition of $\mathbf{p}_i$ as the center of $M$ positive embeddings. This definition represents a deterministic event, and hence its probability distribution $Q(z_i)$ is determined to be 1. Eq.\eqref{eq:M3} arises from the fixed latent variable $\mathbf{p}_i$, as BPR models the likelihood of a user preferring positive examples over negative examples using a sigmoid function.

As can be seen, the feature augmentation in Eq.~\eqref{eq:featureaug}, which in fact performs the expectation step with latent variables interpreted as the interest center for positive examples. And optimization in Eq.~\eqref{eq:theta}, which in fact performs the maximization step. It is worth noting that the E-step differs from previous research~\cite{Li:2021:ICLR,lin2022improving} where latent variables represented the category centers of all items clusters. The expectation step performed in this method not only avoids the computational overhead introduced by clustering all items but also has a meaningful interpretation: the representation of the user's interest center.

\begin{lemma}\label{lemma:2}
	Denote the relative ranking position of item $j$ as $r_{uj} = \frac{1}{|\mathcal{I}|} \sum_{l\in \mathcal{I}}\mathbb{I}_{|\hat{x}_{ul} \leq \hat{x}_{uj}|}$, where $r_{uj}=1$ corresponds to the highest ranking and $r_{uj}= \frac{1}{|\mathcal{I}|}$ corresponds to the lowest ranking. Then, the probability of item $j$ being labeled as a negative example for model training by Eq.~\eqref{eq:la1} and Eq.~\eqref{eq:la2} is:
	\begin{eqnarray} 
		\mathbb{P}(j) 
		&=& \frac{(1-\alpha)}{C_{|\mathcal{I}|}^2} +\frac{(2\alpha-1)}{C_{|\mathcal{I}|}^2}  r_{uj} \nonumber
	\end{eqnarray}
\end{lemma}
\textbf{Proof}: When randomly and uniformly sampling two items from item set $\mathcal{I}$, the probability of item $j$ being included is given by:
\begin{eqnarray}
	\mathbb{P}_1 = \frac{1}{C_{|\mathcal{I}|}^2}
\end{eqnarray}
Furthermore, based on Eq.~\eqref{eq:la1} and Eq.~\eqref{eq:la2}, if item $j$ is labeled as a negative example, there are two cases to consider. The first case is when the similarity score of item $j$ is relatively high, and it is labeled as a negative example with a probability of $\alpha$ via Eq.~\eqref{eq:la1}. The corresponding probability for this case is given by:
\begin{eqnarray}
	\mathbb{P}_2 &=& \mathbb{P}(\hat{x}_{uj} >\hat{x}_{uj^\prime} )\cdot \alpha \\
	&=& \alpha \cdot r_{uj}
\end{eqnarray}
This equation holds because the probability of the event $\hat{x}_{uj} >\hat{x}_{uj^\prime}$ occurring is equivalent to $j^\prime$ having a ranking position within the interval $[0, r_{uj}]$. For uniformly sampled $j^\prime$, this probability is equal to $r_{uj}$.

The second case is when the similarity score of item $j$ is relatively low, but it is still labeled as negative example with a probability of $1-\alpha$ via Eq.~\eqref{eq:la2}. The corresponding probability is given by:
\begin{eqnarray}
	\mathbb{P}_2 &=& \mathbb{P}(\hat{x}_{uj} \leq \hat{x}_{uj^\prime} )\cdot (1-\alpha) \\
	&=& (1-\alpha) \cdot (1-r_{uj})
\end{eqnarray}
Therefore, the probability of item $j$ being labeled as negative sample for model training is given by:
\begin{eqnarray}
	\mathbb{P}(j) &=& \mathbb{P}_1 \cdot (\mathbb{P}_2 + \mathbb{P}_3) \\
	&=& \frac{1}{C_{|\mathcal{I}|}^2} (1-\alpha - r_{uj} + 2\alpha \cdot r_{uj}) \\
	&=& \frac{(1-\alpha)}{C_{|\mathcal{I}|}^2} +\frac{(2\alpha-1)}{C_{|\mathcal{I}|}^2}  r_{uj}
\end{eqnarray}
This implies that when $\alpha > 0.5$, the probability of any item $j$ being labeled as negative sample for model training is proportional to its relative ranking position. The higher the ranking position of $j$ , the greater the probability of $j$ being sampled. Consequently, this approach achieves hard negative sampling with constant time complexity.

\section{Experiment}
\subsection{Experiment Settings}
\subsubsection{Dataset}
Our experimental evaluation is conducted on five publicly available datasets, namely MovieLens-100k, MovieLens-1M, Yahoo!-R3, Yelp2018, and Gowalla. To ensure consistency, we follow existing methods~\cite{Steffen:2009:UAI,Zhang:2013:SIGIR,Steffen:2014:WSDM} and convert all rated items to implicit feedback across the first three datasets. The data is partitioned randomly, with 20\% reserved for testing purposes and the remaining 80\% utilized for training. Table~\ref{Table:Dataset} provides an overview of the dataset statistics, capturing essential details regarding the datasets employed in our study.
\begin{table*}[h!]
	\centering
	\small
	\caption{Dataset Statistics}\label{Table:Dataset}
	\begin{tabular}{lrrrrrr}
		\toprule[1.2pt]
		Dataset          & Users   & Items  &Interactions & Training set  &Test set&Density  \\ \cline{1-7}
		MovieLens-100k   &   943    &  1,682   &100,000&    80k	   & 20k &0.06304\\
		MovieLens-1M    &   6,040  &  3,952   &1,000,000&  800k     & 200k&0.04189  \\
		Yahoo!-R3       &   5,400  &  1,000  &182,000 &   146k      & 36k&0.03370\\
		Yelp2018       &   31,668  &  38,048&   1,561,406&   1,249k     & 312k&0.00130  \\
		Gowalla       &   29,858 &  40,981  &1,027,370 &   821k     & 205k&0.00084 \\
		\bottomrule[1.2pt]
	\end{tabular}
\end{table*}
\subsubsection{Evaluation metric}
To assess the quality of our recommendations, we employ well-established evaluation metrics, namely precision (P), recall (R), and normalized discounted cumulative gain (NDCG). These metrics are commonly used in the field and provide valuable insights into the effectiveness of our top-$K$ recommendations, with $K$ being set to 5, 10, and 20. Given the familiarity of these metrics within the research community, we refrain from reiterating their definitions in this context to maintain conciseness.

\subsubsection{Baselines}
In this study, we modify the widely adopted self-supervised signal of "user prefers  interacted items over non-interacted items" to "user preference for the most representative interacted items over preference for difficult negative examples". This modification effectively changes the optimization objective. Based on this, we primarily compare our approach with two categories of methods: ranking oriented loss function modification methods, which aim to modify the commonly used contrastive loss under the self-supervised setting, and self-supervised recommendation methods.

\begin{table*}[!]
	\centering
	\caption{Top-k ranking performance compared with different optimization criterion.}\label{5Table:Recommendation}
	\resizebox{1\textwidth}{!}{
		\begin{tabular}{lllccccccccccc}
			\toprule[1.2pt]
			\multirow{2}*{\textbf{Datasets}} & \multirow{2}*{\textbf{Encoder}} & \multirow{2}*{\textbf{Optimization Criterion}} & \multicolumn{3}{c}{Top-5} &~& \multicolumn{3}{c}{Top-10}&~&\multicolumn{3}{c}{Top-20}\\ \cline{4-6} \cline{8-10} \cline{12-14}
			~ & ~ & ~ & Precision& Recall& NDCG& ~ &Precision& Recall& NDCG& ~ &Precision& Recall& NDCG \\ \hline
			
			\multirow{10}*{\textbf{MovieLens-100k}} & \multirow{5}*{\textbf{MF}} & BPR & 0.3900   &0.1301	&0.4143	&~&0.3363	&0.2164	&0.3967& ~&0.2724&0.3298&0.3962 \\
			~ & ~ &InfoNCE  &0.4081 & 0.1388 & 0.4324 & ~ & 0.3452 & 0.2266 & 0.4095 & ~ & 0.2793 & 0.3497 & 0.4118 \\
			~ & ~ & DCL &0.4168 & 0.1434 & 0.4458 & ~ & 0.3513 & 0.2291 & 0.4202 & ~ & 0.2835 & 0.3546 & 0.4207 \\ 
			~ & ~ & HCL  & 0.4263 & 0.1463 & 0.4539 & ~ & 0.3565 & 0.2323 & 0.426 & ~ & 0.2849 & 0.3564 & 0.4242 \\	
			~ & ~ &Proposed    &\textbf{0.4380} & \textbf{0.1524} & \textbf{0.4662} & ~ & \textbf{0.3646} & \textbf{0.2367} & \textbf{0.4366} & ~ & \textbf{0.2909} & \textbf{0.3588} & \textbf{0.4357} \\ 
			\cline{2-14}
			~ & \multirow{5}*{\textbf{LightGCN}}  & BPR & 0.3944 & 0.1231 & 0.4204 & ~ & 0.3346 & 0.2189 & 0.4017 & ~ & 0.2658 & 0.3281 & 0.3986 \\ 
			~ & ~ & InfoNCE & 0.3924 & 0.1343 & 0.4209 & ~ & 0.3349 & 0.2183 & 0.4006 & ~ & 0.2679 & 0.3289 & 0.3976 \\ 
			~ & ~ & DCL & 0.3962 & 0.1367 & 0.4243 & ~ & 0.3361 & 0.2194 & 0.4022 & ~ & 0.2695 & 0.3329 & 0.4006 \\ 
			~ & ~ & HCL & 0.4197 & 0.1461 & 0.4501 & ~ & 0.3458 & 0.2256 & 0.4188 & ~ & 0.2802 & 0.3446 & 0.4182 \\
			~ & ~ & Proposed & \textbf{0.4387} & \textbf{0.1532} & \textbf{0.4661} & ~ & \textbf{0.3633} & \textbf{0.2376} & \textbf{0.4362} & ~ & \textbf{0.2947} & \textbf{0.3591} & \textbf{0.4355} \\ \hline\hline
			
			\multirow{10}*{\textbf{MovieLens-1M}} & \multirow{5}*{\textbf{MF}} & BPR & 0.3843    &0.0855	&0.4027	&~&0.3353	&0.1430	&0.3737& ~&0.2798&0.2244&0.3572 \\ 
			~ & ~ & InfoNCE & 0.3820 & 0.0879 & 0.4003 & ~ & 0.3339 & 0.1478 & 0.3728 & ~ & 0.2821 & 0.2358 & 0.3605 \\ 
			~ & ~ & DCL &  0.4009 & 0.0934 & 0.4209 & ~ & 0.3472 & 0.1546 & 0.3894 & ~ & 0.289 & 0.2423 & 0.3731\\ 
			~ & ~ & HCL & 0.4112 & 0.0969 & 0.4317 & ~ & 0.3552 & 0.1585 & 0.3991 & ~ & 0.2959 & 0.2475 & 0.3825 \\ 
			~ & ~ & Proposed & \textbf{0.4229} & \textbf{0.1022} & \textbf{0.4418} & ~ & \textbf{0.3645 }&\textbf{ 0.1644} & \textbf{0.4096} & ~ & \textbf{0.3002} & \textbf{0.2526} & \textbf{0.3912} \\
			\cline{2-14}
			~ & \multirow{5}*{\textbf{LightGCN}} & BPR &0.4095&0.0953&0.4305&~&0.3512&0.1547&0.3985& ~&0.2915&0.2405&0.3781 \\ 
			~ & ~ & InfoNCE & 0.4101 & 0.0986 & 0.4386 & ~ & 0.359 & 0.1594 & 0.4041 & ~ & 0.2979 & 0.2482 & 0.3869 \\ 
			~ & ~ & DCL & 0.4104 & 0.0982 & 0.4291 & ~ & 0.3544 & 0.1597 & 0.3977 & ~ & 0.2965 & 0.2511 & 0.3842 \\ 
			~ & ~ & HCL & 0.4107 & 0.0948 & 0.4300 & ~ & 0.3514 & 0.1542 & 0.3950 & ~ & 0.2916 & 0.2413 & 0.3775 \\ 
			~ & ~ & Proposed & \textbf{0.4256} &\textbf{ 0.1047} &\textbf{ 0.4460} & ~ & \textbf{0.3651 }& \textbf{0.1658} & \textbf{0.1893} & ~ & \textbf{0.2998} &\textbf{ 0.2534}& \textbf{0.3911} \\\hline \hline
			
			\multirow{10}*{\textbf{Yahoo!-R3}} & \multirow{5}*{\textbf{MF}} & BPR & 0.1417 & 0.1052 & 0.1587 & ~ & 0.1064 & 0.1573 & 0.1641 & ~ & 0.0768 & 0.2259 & 0.1913 \\ 
			~ & ~ &  InfoNCE & 0.1429 & 0.1065 & 0.1615 & ~ & 0.1080 & 0.1601 & 0.1664 & ~ & 0.0786 & 0.2316 & 0.1952 \\ 
			~ & ~ & DCL &0.1454 & 0.1083 & 0.1635 & ~ & 0.1091 & 0.1618 & 0.1692 & ~ & 0.079 & 0.2327 & 0.1974  \\ 
			~ & ~ & HCL & 0.1460 & 0.1079 & 0.1638 & ~ & 0.1096 & 0.1628 & 0.1697 & ~ & 0.0792 & 0.2336 & 0.1976 \\ 
			~ & ~ & Proposed & \textbf{0.1529}& \textbf{0.1122} & \textbf{0.1685} & ~ & \textbf{0.1124 }& \textbf{0.1667} & \textbf{0.1736} & ~ & \textbf{0.0815} & \textbf{0.2367} & \textbf{0.2029} \\
			\cline{2-14}
			~ & \multirow{5}*{\textbf{LightGCN}} &BPR & 0.1479&0.1101&0.1693&~&0.1126&0.1669&0.1760&~& 0.0814&0.2389&0.2047 \\ 
			~ & ~ & InfoNCE & 0.1417 & 0.1074 & 0.1676 & ~ & 0.1099 & 0.1633 & 0.1719 & ~ & 0.0798 & 0.2354 & 0.2007  \\ 
			~ & ~ & DCL & 0.1456 & 0.1092 & 0.1642 & ~ & 0.1089 & 0.1622 & 0.1697 & ~ & 0.079 & 0.2333 & 0.1982\\ 
			~ & ~ & HCL & 0.1512 & 0.1119 & 0.1718 & ~ & 0.1130 & 0.1683 & 0.1766 & ~ & 0.0812 & 0.2394 & 0.2049 \\ 
			~ & ~ & Proposed & \textbf{0.1537} & \textbf{0.1140} &  \textbf{0.1725} & ~ & \textbf{0.1156} & \textbf{0.1696}  & \textbf{0.1771} & ~ & \textbf{0.0842} & \textbf{0.2428} & \textbf{0.2076}\\ \hline\hline
			
			\multirow{10}*{\textbf{Yelp2018}} & \multirow{5}*{\textbf{MF}} & BPR & 0.0398 & 0.0228 & 0.0435 & ~ & 0.0339 & 0.0389 & 0.0456 & ~ & 0.0284 & 0.065 & 0.0538 \\ 
			~ & ~ &  InfoNCE  & 0.0429 & 0.0246 & 0.047 & ~ & 0.0365 & 0.0417 & 0.0491 & ~ & 0.0305 & 0.07 & 0.058 \\ 
			~ & ~ & DCL & 0.0486 & 0.0278 & 0.0531 & ~ & 0.0410 & 0.0466 & 0.0552 & ~ & 0.0342 & 0.0777 & 0.0648 \\
			~ & ~ & HCL & 0.0515 & 0.0305 & 0.0566 & ~ & 0.0459 & 0.0541 & 0.0622 & ~ & 0.0383 & 0.0894& 0.0736 \\ 
			~ & ~ & Proposed & \textbf{0.0586} &\textbf{ 0.0378} & \textbf{0.0633} & ~ & \textbf{0.0481} &\textbf{ 0.0592} & \textbf{0.0671} & ~ & \textbf{0.0412} & \textbf{0.0950} & \textbf{0.0783} \\
			\cline{2-14}
			~ & \multirow{5}*{\textbf{LightGCN}} & BPR & 0.0556 & 0.0330 & 0.0610 & ~ & 0.0473 & 0.0560 & 0.0644 & ~ & 0.0391 & 0.0914 & 0.0757 \\ 
			~ & ~ & InfoNCE & 0.0553 & 0.0329 & 0.0607 & ~ & 0.0473 & 0.0558 & 0.0642 & ~ & 0.0390 & 0.0911 & 0.0754 \\ 
			~ & ~ & DCL & 0.0559 & 0.0331 & 0.0612 & ~ & 0.0472 & 0.0557 & 0.0642 & ~ & 	0.0391 & 0.0914 & 0.0756 \\ 
			~ & ~ & HCL & 0.0563 & 0.0335 & 0.0617 & ~ & 0.0477 & 0.0564 & 0.0648 & ~ & 0.0393 & 0.0920 & 0.0760 \\  
			~ & ~ & Proposed & \textbf{0.0645} & \textbf{0.0401} & \textbf{0.0702} & ~ & \textbf{0.0555} & \textbf{0.0658} &\textbf{ 0.0732} & ~ & \textbf{0.0446} & \textbf{0.1047} & \textbf{0.0849} \\\hline\hline
			
			\multirow{10}*{\textbf{Gowalla}} & \multirow{5}*{\textbf{MF}} & BPR & 0.0728 & 0.0748 & 0.1000 & ~ & 0.0555 & 0.1116 & 0.1063 & ~ & 0.0414 & 0.1625 & 0.1209 \\ 
			~ & ~  & InfoNCE & 0.0739 &0.0757 & 0.1016 & ~ & 0.0560 & 0.1122 & 0.1076 & ~ & 0.0422 & 0.1650 & 0.1230\\ 
			~ & ~ & DCL & 0.0746 & 0.0769 &0.1023 & ~ & 0.0568 & 0.1147 & 0.1088 & ~ & 0.0426 & 0.1664 & 0.1238 \\ 
			~ & ~ & HCL & 0.0755 & 0.0774 & 0.1035 & ~ & 0.0574 & 0.1151 & 0.1098 & ~ & 0.0432& 0.1693 & 0.1256 \\ 
			~ & ~ & Proposed & \textbf{0.0832} & \textbf{0.0841} & \textbf{0.1130} & ~ & \textbf{0.0645} & \textbf{0.1263}& \textbf{0.1190} & ~ & \textbf{0.0488} & \textbf{0.1827} & \textbf{0.1356} \\
			\cline{2-14}
			~ & \multirow{5}*{\textbf{LightGCN}} & BPR & 0.0735 & 0.0753 & 0.1007 & ~ & 0.0560 & 0.1119 & 0.1069 & ~ & 0.0419 & 0.1641 & 0.1218 \\ 
			~ & ~ & InfoNCE & 0.0743 & 0.0760 & 0.1022 & ~ & 0.0566 & 0.1132& 0.1084 & ~ & 0.0423 & 0.1649 & 0.1231\\ 
			~ & ~ & DCL & 0.0748 & 0.0763 & 0.1027 & ~ & 0.0569 & 0.1132 & 0.1088 & ~ & 0.0424 & 0.1656 & 0.1236 \\ 
			~ & ~ & HCL & 0.0852 & 0.0874 & 0.1156 & ~ & 0.0658 &0.1321 & 0.1237 & ~ & 0.0495 & 0.1930 & 0.1409 \\ 
			~ & ~ & Proposed & \textbf{0.0896} & \textbf{ 0.0926} & \textbf{0.1207} & ~ &  \textbf{0.0692} &  \textbf{0.1375} & \textbf{0.1290} & ~ & \textbf{0.0504} &\textbf{ 0.1991} & \textbf{0.1478}  \\\hline
			\bottomrule[1.5pt]
			
		\end{tabular}
	}
\end{table*}
\begin{itemize}
	\item Loss function modification methods:
	\subitem BPR~\cite{Steffen:2009:UAI}(UAI,2008): BPR introduces the pairwise learning approach based on maximum a posteriori estimation for implicit collaborative filtering. Its  meaning is introduced in Eq.~\eqref{eq:bpr}.
	
	\subitem InfoNCE~\cite{Oord:2018:arxiv}(arXiv,2018):
	The InfoNCE loss is a popular loss function used in machine learning, particularly in the context of representation learning.  Specifically, InfoNCE measures the similarity between a query sample and the set of negative samples. InfoNCE can be seen as a generalization of BPR from one negative sample to $N$ negative samples. In practice, since label of negative samples are unavailable, negative samples are typically  sampled from unlabeled samples. Its  meaning is introduced in Eq.~\eqref{eq:infonce}.
	
	\subitem DCL~\cite{Chuang:2020:NIPS}(NeurIPS spotlight,2020): Due to the unavailability of negative labels in the self-supervised setting, DCL corrects the probability estimates to perform false negative debiasing. Specifically, it proposes the estimator to replace the second term in the denominator of the $\mathcal{L}_\text{InfoNCE}$:
	\begin{eqnarray}\label{eq:DCL}
		\mathcal{L}_\text{DCL}
		&=&  - \mathbb{E}\log\frac{\exp(g(\hat{x}_{ui})}{\exp(\hat{x}_{ui})+ N\phi}
	\end{eqnarray}
	where
	\begin{eqnarray}
		\phi =  \frac{1}{N\tau^-}  (\sum_{j=1}^{N} \exp(\hat{x}_{uj} - N\tau^+ \cdot \frac{\sum_{i=1}^{K} \exp(\hat{x}_{ui})}{K} ) \label{Eq:DCLEstimator}
	\end{eqnarray}
	The $\phi$ can be interpreted as the estimation of expectation of similarity scores for true negative samples.
	
	\item HCL~\cite{Robinson:2021:ICLR}(ICLR,2021): Following the DCL debiasing framework, it also takes into consideration of hard negative mining by up-weighting each randomly selected unlabeled sample as follows.
	\begin{eqnarray}\label{eq:hcl}
		\omega_i^\textsc{Hcl} = \frac{\hat{x}_{uj}^\beta}{\frac{1}{N} \sum_{j=1}^{N}\hat{x}_{uj}^\beta}.
	\end{eqnarray}
	where beta controls the hardness level for mining hard negatives. DCL is a particular case of HCL with $\beta=0$.
	\item Self-supervised recommendation methods:
	\subitem SGL~\cite{10.1145/3404835.3462862}(SIGIR,2021): The idea is to enhance the traditional supervised recommendation task by incorporating an auxiliary self-supervised task that strengthens node representation learning through graph contrastive learning. In this approach, we apply node dropout, aiming to maximize the agreement between different views of the same node while differentiating them from other nodes.
	\subitem NCL~\cite{lin2022improving}(WWW,2022): The idea of neighborhood enriched Contrastive Learning explicitly integrates potential neighbors into contrastive pairs. Specifically, it incorporates the neighbors of a user (or an item) from both the graph structure and semantic space.
	\subitem SimGCL~\cite{lin2022improving}(SIGIR,2022): The core idea of SimGCL is to directly inject noise into the embeddings in the embedding space to enhance the feature representation. This aims to obtain relatively smooth feature representations and thereby reduce popularity bias to improve ranking accuracy.
\end{itemize}

\subsubsection{Experimental Setup}

For loss function correction methods, two representative backbone models are employed  to validate the effectiveness of our modified self-supervised signal: the matrix factorization (MF)\cite{Koren:2009:Computer} and  (LightGCN)\cite{Xiangnan:2020:SIGIR}. For self-supervised recommendation methods, as they all propose their own encoders, we fix the encoder of this method as MF for comparison. The computations for the first three datasets were conducted on a personal computer running the Windows 10 operating system with a 4.1 GHz CPU, an RTX 3060 GPU, and 32 GB of RAM. The computations for the last two datasets were performed on a cloud server running the Linux operating system with a Xeon(R) Platinum 8358P CPU, an RTX A40 GPU, and 56GB of RAM. The code and corresponding parameters have been released at: \url{https://github.com/} for reproducibility.

\subsection{Top-k Ranking Performance}

From Table.~\ref{5Table:Recommendation}, it can be observed that our proposed method achieves the best performance across different datasets and recommendation models. Compared to BPR and InfoNCE without debiasing mechanisms, our method exhibits significant improvements, but maintains the identical running time, highlighting the necessity of enhancing the original self-supervised signals. Additionally, DCL and HCL, which include debiasing mechanisms, generally outperform BPR and InfoNCE without such mechanisms. In comparison to DCL and HCL with debiasing mechanisms, our method also achieves performance improvements. This is mainly attributed to the comprehensive consideration of the issues related to pseudo-positive and pseudo-negative samples, which enhances the self-supervised signals. In contrast, DCL and HCL primarily focus on improving the supervised signals for negative samples.

The second observation is that, on large-scale sparse datasets, BPR (a special case of InfoNCE with only one negative sample) generally performs worse than InfoNCE with multiple negative samples. This is because a larger number of negative samples $N$  lower bounds the mutual information~\cite{Oord:2018:arxiv}, resulting in performance improvements. Moreover, a larger $N$ typically assigns larger gradient values to hard samples, thereby benefiting from hard negative mining~\cite{Jiancan:2022:arxiv}. Specifically, HCL assigns higher weights to difficult samples based on their high scores, which leads to larger gradient values for these samples and enables the mining of hard samples. This approach often achieves suboptimal results, particularly on sparse datasets, indicating the importance of hard negative mining when the probability of false-negative samples is low.

As shown in Table.~\ref{6Table:Recommendation}, SGL performs relatively poorly, primarily because its self-supervised method is primarily designed for graph contrast tasks. However, graph contrast tasks primarily aim to learn graph structural information, which is inconsistent with ranking tasks. The main role of graph contrast is to smooth the representations, and the effectiveness of self-supervised tasks specifically designed for graphs contrast in personalized ranking tasks is questionable~\cite{lin2022improving}. Furthermore, it requires joint optimization with the ranking objective, which poses challenges in coordinating multiple self-supervised tasks and often involves manual hyperparameter search to balance different self-supervised tasks. NCL shares a similar idea to our method in handling positive noise; it obtains prototypes through sample clustering for graph contrast rather than preference contrast, thus making limited contributions to ranking tasks. The most competitive method is SimGCL, which is also the most computationally efficient approach. While it achieves better performance through feature smoothing, it does not provide specific improvements for the ranking oriented task.

\begin{table*}[!h]
	\centering
	\caption{Top-k ranking performance compared with different self-supervised recommendation methods.}\label{6Table:Recommendation}
	\resizebox{1\textwidth}{!}{
		\begin{tabular}{lllccccccccccc}
			\toprule[1.2pt]
			\multirow{2}*{\textbf{Datasets}} & \multirow{2}*{\textbf{~}} & \multirow{2}*{\textbf{SSR methods}} & \multicolumn{3}{c}{Top-5} &~& \multicolumn{3}{c}{Top-10}&~&\multicolumn{3}{c}{Top-20}\\ \cline{4-6} \cline{8-10} \cline{12-14}
			~ & ~ & ~ & Precision& Recall& NDCG& ~ &Precision& Recall& NDCG& ~ &Precision& Recall& NDCG \\ \hline
			
			\multirow{4}*{\textbf{MovieLens-100k}} & \multirow{5}*{~} & SGL &0.4091 & 0.1395 & 0.4333 & ~ & 0.3458 & 0.2276 & 0.4096 & ~ & 0.2791 & 0.3498 & 0.4120 \\
			~ & ~ &NCL  &0.4181 & 0.1482 & 0.4404 & ~ & 0.3522 & 0.2306 & 0.4212 & ~ & 0.2852 & 0.3542 & 0.4258 \\
			~ & ~ & SimGCL  & 0.4303 & 0.1503 & 0.4619 & ~ & 0.3615 & 0.2323 & 0.4326 & ~ & 0.2849 & 0.3564 & 0.4242 \\	
			~ & ~ &Proposed    &\textbf{0.4380} & \textbf{0.1524} & \textbf{0.4662} & ~ & \textbf{0.3646} & \textbf{0.2367} & \textbf{0.4366} & ~ & \textbf{0.2909} & \textbf{0.3588} & \textbf{0.4357} \\ 
			\cline{2-14}
			
			\multirow{4}*{\textbf{MovieLens-1M}} & \multirow{5}*{~} & SGL &  0.4012 & 0.0942 & 0.4218 & ~ & 0.3477 & 0.1542 & 0.3904 & ~ & 0.2895 & 0.2432 & 0.3735\\ 
			~ & ~ & NCL &  0.4102 & 0.1024 & 0.4311 & ~ & 0.3551 & 0.1623 & 0.3971 & ~ & 0.2951 & 0.2480 & 0.3809\\ 
			~ & ~ & SimGCL & 0.4172 & 0.0977 & 0.4382 & ~ & 0.3595 & 0.1605 & 0.4056 & ~ & 0.2963 & 0.2481 & 0.3877 \\ 
			~ & ~ & Proposed & \textbf{0.4229} & \textbf{0.1022} & \textbf{0.4418} & ~ & \textbf{0.3645 }&\textbf{ 0.1644} & \textbf{0.4096} & ~ & \textbf{0.3002} & \textbf{0.2526} & \textbf{0.3912} \\
			\cline{2-14}
			
			\multirow{4}*{\textbf{Yahoo!-R3}} & \multirow{5}*{~} & SGL & 0.1464 & 0.1085 & 0.1637 & ~ & 0.1085 & 0.1623 & 0.1695 & ~ & 0.0801 & 0.2322 & 0.1978 \\
			~ & ~ & NCL &0.1461 & 0.1088 & 0.1642 & ~ & 0.1099 & 0.1625 & 0.1697 & ~ & 0.0788 & 0.2315 & 0.1983  \\ 
			~ & ~ & SimGCL & 0.1488 & 0.1095 & 0.1658 & ~ & 0.1090 & 0.1632 & 0.1691 & ~ & 0.0798 & 0.2331 & 0.1997 \\ 
			~ & ~ & Proposed & \textbf{0.1529}& \textbf{0.1122} & \textbf{0.1685} & ~ & \textbf{0.1124 }& \textbf{0.1667} & \textbf{0.1736} & ~ & \textbf{0.0815} & \textbf{0.2367} & \textbf{0.2029} \\
			\cline{2-14}
			
			\multirow{4}*{\textbf{Yelp2018}} & \multirow{5}*{~} & SGL & 0.0513 & 0.0308 & 0.0569 & ~ & 0.0461 & 0.0543 & 0.0621 & ~ & 0.0380 & 0.0896& 0.0735 \\ 
			~ & ~ & NCL & 0.0541 & 0.0332 & 0.0592 & ~ & 0.0471 & 0.0550 & 0.0639 & ~ & 0.0382 & 0.0899& 0.0712 \\ 
			~ & ~ & SimGCL & 0.0560 & 0.0340 & 0.0599 & ~ & 0.0487 & 0.0561 & 0.0651 & ~ & 0.0394 & 0.0932& 0.0763 \\
			~ & ~ & Proposed & \textbf{0.0586} &\textbf{ 0.0378} & \textbf{0.0633} & ~ & \textbf{0.0481} &\textbf{ 0.0592} & \textbf{0.0671} & ~ & \textbf{0.0412} & \textbf{0.0950} & \textbf{0.0783} \\
			\cline{2-14}
			
			\multirow{4}*{\textbf{Gowalla}} & \multirow{5}*{~} & SGL & 0.0740 & 0.0763 &0.1021 & ~ & 0.0566 & 0.1145 & 0.1086 & ~ & 0.0422 & 0.1664 & 0.1236 \\  
			~ & ~ & NCL & 0.0755 & 0.0778 &0.1032 & ~ & 0.0573 & 0.1152 & 0.1093 & ~ & 0.0428 & 0.1667 & 0.1241 \\ 
			~ & ~ & SimGCL & 0.0785 & 0.0791 & 0.1082 & ~ & 0.0593 & 0.1212 & 0.1142 & ~ & 0.0443& 0.1753 & 0.1302 \\ 
			~ & ~ & Proposed & \textbf{0.0832} & \textbf{0.0841} & \textbf{0.1130} & ~ & \textbf{0.0645} & \textbf{0.1263}& \textbf{0.1190} & ~ & \textbf{0.0488} & \textbf{0.1827} & \textbf{0.1356} \\
			\cline{2-14}
			\bottomrule[1.5pt]
			
		\end{tabular}
	}
\end{table*}

\subsection{Hyperparameter Analysis}
This method involves only two hyperparameters: $M$, which controls the number of positive samples for feature augmentation, and $\alpha$, which controls the probability of sampling hard negative samples. On the MovieLens100K dataset, we first fix $\alpha=1$ and examine the impact of different values of $M$ on the precision@5 metric.
\begin{figure}[!h]
	\centering
	\includegraphics[width=0.49\textwidth]{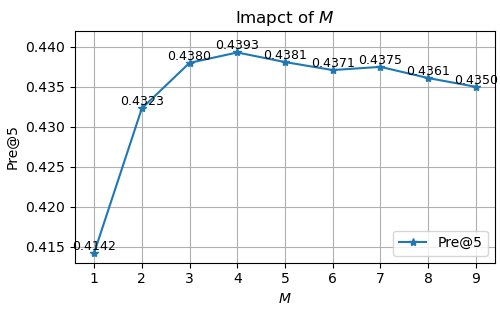}
	\caption{Impact of different values of $M$.}
	\label{Fig:M}
\end{figure}
When $M=1$, positive feature augmentation has no effect, and the performance improvement relative to BPR is mainly attributed to the label augmentation of negative samples. When $M>1$, the top-5 accuracy begins to improve, reaching its peak at $M=4$. Subsequently, as $M$ increases further, the top-5 accuracy starts to slowly decline. We hypothesize that this may be due to an excessive number of positive examples used for feature augmentation, resulting in excessive smoothing of the embeddings. In practice, users often have multiple interests, and sampling a smaller number of $M$ at each training epoch helps to learn the various interests of the users.
\begin{figure}[h!]
	\centering
	\includegraphics[width=0.49\textwidth]{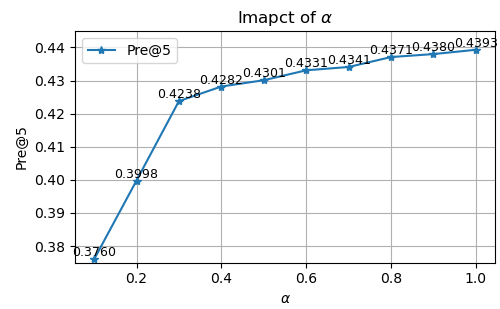}
	\caption{Impact of different values of $\alpha$.}
	\label{Fig:A}
\end{figure}

Figure~\ref{Fig:A} illustrates the impact of different values of $\alpha$ on top-5 accuracy when fixing $M=4$. Notably, when $\alpha$ is set to 0.5, it corresponds to uniform random sampling, as demonstrated in Lemma~\ref{lemma:2}. In this case, the performance improvement compared to BPRMF is primarily attributed to the positive feature augmentation. As $\alpha$ increases beyond 0.5, the probability of sampling hard negative samples also increases, resulting in consistent performance enhancements. This observation supports the notion that items exposed to users but not interacted with reflect negative preferences. Consequently, in practical applications, it is advisable to select an $\alpha$ value between 0.5 and 1. Since the top-5 accuracy is 0.43 for $\alpha$=0.5, out of the 4.8 percentage point improvement achieved by this method relative to BPRMF, 4 percentage points are attributed to positive feature augmentation, and 0.8 percentage points are attributed to negative feature augmentation.

\section{Conclusion}
Contrastive learning-based recommendation algorithms have significantly advanced the field of self-supervised recommendation, with BPR serving as a prominent example of preference contrast tasks. These algorithms optimize the self-supervisory signal by ensuring that positive scores surpass negative scores, thereby optimizing the AUC and NDCG of ranking lists and dominating implicit collaborative filtering tasks. However, the presence of false-positive and false-negative examples in recommendation systems hampers accurate preference learning.

In this study, we propose a simple self-supervised contrastive learning framework that leverages positive feature augmentation and negative label augmentation to improve the self-supervisory signal. Theoretically, we demonstrate that our learning method is equivalent to maximizing the likelihood estimation with latent variables representing user interest centers. Additionally, we establish that our negative label augmentation technique samples unlabeled examples with a probability linearly dependent on their relative ranking positions, enabling efficient augmentation in constant time complexity. Through validation on multiple datasets, we illustrate the significant improvements our method achieves over the widely used BPR optimization objective while maintaining comparable runtime. Our future work involves leveraging commonly available textual and image information to further optimize the self-supervisory signal proposed in this study, thereby enhancing the accuracy of recommendation systems.
\normalem
\bibliographystyle{ACM-Reference-Format}
\bibliography{refs}


\begin{thebibliography}{48}


\ifx \showCODEN    \undefined \def \showCODEN     #1{\unskip}     \fi
\ifx \showDOI      \undefined \def \showDOI       #1{#1}\fi
\ifx \showISBNx    \undefined \def \showISBNx     #1{\unskip}     \fi
\ifx \showISBNxiii \undefined \def \showISBNxiii  #1{\unskip}     \fi
\ifx \showISSN     \undefined \def \showISSN      #1{\unskip}     \fi
\ifx \showLCCN     \undefined \def \showLCCN      #1{\unskip}     \fi
\ifx \shownote     \undefined \def \shownote      #1{#1}          \fi
\ifx \showarticletitle \undefined \def \showarticletitle #1{#1}   \fi
\ifx \showURL      \undefined \def \showURL       {\relax}        \fi
\providecommand\bibfield[2]{#2}
\providecommand\bibinfo[2]{#2}
\providecommand\natexlab[1]{#1}
\providecommand\showeprint[2][]{arXiv:#2}

\bibitem[Bachman et~al\mbox{.}(2019)]%
        {Bachman:2019:NIPS}
\bibfield{author}{\bibinfo{person}{Philip Bachman}, \bibinfo{person}{R~Devon
  Hjelm}, {and} \bibinfo{person}{William Buchwalter}.}
  \bibinfo{year}{2019}\natexlab{}.
\newblock \showarticletitle{Learning representations by maximizing mutual
  information across views}.
\newblock  (\bibinfo{year}{2019}).
\newblock


\bibitem[Bian et~al\mbox{.}(2021)]%
        {bian2021contrastive}
\bibfield{author}{\bibinfo{person}{Shuqing Bian}, \bibinfo{person}{Wayne~Xin
  Zhao}, \bibinfo{person}{Kun Zhou}, \bibinfo{person}{Jing Cai},
  \bibinfo{person}{Yancheng He}, \bibinfo{person}{Cunxiang Yin}, {and}
  \bibinfo{person}{Ji-Rong Wen}.} \bibinfo{year}{2021}\natexlab{}.
\newblock \showarticletitle{Contrastive curriculum learning for sequential user
  behavior modeling via data augmentation}. In
  \bibinfo{booktitle}{\emph{Proceedings of the 30th ACM International
  Conference on Information \& Knowledge Management}}.
  \bibinfo{pages}{3737--3746}.
\newblock


\bibitem[Cai et~al\mbox{.}(2023)]%
        {lightgcl:2023:ICLR}
\bibfield{author}{\bibinfo{person}{Xuheng Cai}, \bibinfo{person}{Chao Huang},
  \bibinfo{person}{Lianghao Xia}, {and} \bibinfo{person}{Xubin Ren}.}
  \bibinfo{year}{2023}\natexlab{}.
\newblock \showarticletitle{LightGCL: Simple Yet Effective Graph Contrastive
  Learning for Recommendation}.
\newblock \bibinfo{journal}{\emph{International Conference on Learning
  Representations}} (\bibinfo{year}{2023}).
\newblock


\bibitem[Chen et~al\mbox{.}(2020)]%
        {Chen:2020:NIPS}
\bibfield{author}{\bibinfo{person}{Ting Chen}, \bibinfo{person}{Simon
  Kornblith}, \bibinfo{person}{Kevin Swersky}, \bibinfo{person}{Mohammad
  Norouzi}, {and} \bibinfo{person}{Geoffrey~E Hinton}.}
  \bibinfo{year}{2020}\natexlab{}.
\newblock \showarticletitle{Big self-supervised models are strong
  semi-supervised learners}.
\newblock \bibinfo{journal}{\emph{Proceedings of the International Conference
  on Neural Information Processing Systems}}  \bibinfo{volume}{33}
  (\bibinfo{year}{2020}), \bibinfo{pages}{22243--22255}.
\newblock


\bibitem[Chu et~al\mbox{.}(2023)]%
        {Chu:2023:PAMI}
\bibfield{author}{\bibinfo{person}{Jielei Chu}, \bibinfo{person}{Jing Liu},
  \bibinfo{person}{Hongjun Wang}, \bibinfo{person}{Hua Meng},
  \bibinfo{person}{Zhiguo Gong}, {and} \bibinfo{person}{Tianrui Li}.}
  \bibinfo{year}{2023}\natexlab{}.
\newblock \showarticletitle{Micro-Supervised Disturbance Learning: A
  Perspective of Representation Probability Distribution}.
\newblock \bibinfo{journal}{\emph{IEEE Transactions on Pattern Analysis and
  Machine Intelligence}}  \bibinfo{volume}{45} (\bibinfo{year}{2023}),
  \bibinfo{pages}{7542--7558}.
\newblock


\bibitem[Chuang et~al\mbox{.}(2020)]%
        {Chuang:2020:NIPS}
\bibfield{author}{\bibinfo{person}{Ching-Yao Chuang}, \bibinfo{person}{Joshua
  Robinson}, \bibinfo{person}{Yen-Chen Lin}, \bibinfo{person}{Antonio
  Torralba}, {and} \bibinfo{person}{Stefanie Jegelka}.}
  \bibinfo{year}{2020}\natexlab{}.
\newblock \showarticletitle{Debiased Contrastive Learning}. In
  \bibinfo{booktitle}{\emph{Proceedings of the International Conference on
  Neural Information Processing Systems}}. \bibinfo{pages}{8765--8775}.
\newblock


\bibitem[Dempster et~al\mbox{.}(1977)]%
        {Dempster:1977:RSS}
\bibfield{author}{\bibinfo{person}{Arthur~P Dempster}, \bibinfo{person}{Nan~M
  Laird}, {and} \bibinfo{person}{Donald~B Rubin}.}
  \bibinfo{year}{1977}\natexlab{}.
\newblock \showarticletitle{Maximum likelihood from incomplete data via the EM
  algorithm}.
\newblock \bibinfo{journal}{\emph{Journal of the Royal Statistical Society:
  Series B (Methodological)}} \bibinfo{volume}{39}, \bibinfo{number}{1}
  (\bibinfo{year}{1977}), \bibinfo{pages}{1--22}.
\newblock


\bibitem[Ding et~al\mbox{.}(2019)]%
        {Jingtao:2019:IJCAI}
\bibfield{author}{\bibinfo{person}{Jingtao Ding}, \bibinfo{person}{Yuhan Quan},
  \bibinfo{person}{Xiangnan He}, \bibinfo{person}{Yong Li}, {and}
  \bibinfo{person}{Depeng Jin}.} \bibinfo{year}{2019}\natexlab{}.
\newblock \showarticletitle{Reinforced Negative Sampling for Recommendation
  with Exposure Data}. In \bibinfo{booktitle}{\emph{Proceedings of the
  Twenty-Eighth International Joint Conference on Artificial Intelligence}}.
  \bibinfo{pages}{2230--2236}.
\newblock


\bibitem[Ding et~al\mbox{.}(2020)]%
        {Ding:2020:NIPS}
\bibfield{author}{\bibinfo{person}{Jingtao Ding}, \bibinfo{person}{Yuhan Quan},
  \bibinfo{person}{Quanming Yao}, \bibinfo{person}{Yong Li}, {and}
  \bibinfo{person}{Depeng Jin}.} \bibinfo{year}{2020}\natexlab{}.
\newblock \showarticletitle{Simplify and Robustify Negative Sampling for
  Implicit Collaborative Filtering}. In \bibinfo{booktitle}{\emph{Proceedings
  of the International Conference on Neural Information Processing Systems}}.
\newblock


\bibitem[Dosovitskiy et~al\mbox{.}(2014)]%
        {Dosovitskiy:2014:NIPS}
\bibfield{author}{\bibinfo{person}{Alexey Dosovitskiy},
  \bibinfo{person}{Jost~Tobias Springenberg}, \bibinfo{person}{Martin
  Riedmiller}, {and} \bibinfo{person}{Thomas Brox}.}
  \bibinfo{year}{2014}\natexlab{}.
\newblock \showarticletitle{Discriminative unsupervised feature learning with
  convolutional neural networks}. In \bibinfo{booktitle}{\emph{Proceedings of
  the International Conference on Neural Information Processing Systems}}.
  \bibinfo{pages}{766–774}.
\newblock


\bibitem[Geng et~al\mbox{.}(2022)]%
        {geng2022recommendation}
\bibfield{author}{\bibinfo{person}{Shijie Geng}, \bibinfo{person}{Shuchang
  Liu}, \bibinfo{person}{Zuohui Fu}, \bibinfo{person}{Yingqiang Ge}, {and}
  \bibinfo{person}{Yongfeng Zhang}.} \bibinfo{year}{2022}\natexlab{}.
\newblock \showarticletitle{Recommendation as language processing (rlp): A
  unified pretrain, personalized prompt \& predict paradigm (p5)}. In
  \bibinfo{booktitle}{\emph{Proceedings of the 16th ACM Conference on
  Recommender Systems}}. \bibinfo{pages}{299--315}.
\newblock


\bibitem[Grill et~al\mbox{.}(2020)]%
        {BYOL:2020:NIPS}
\bibfield{author}{\bibinfo{person}{Jean-Bastien Grill},
  \bibinfo{person}{Florian Strub}, \bibinfo{person}{Florent Altch\'{e}},
  \bibinfo{person}{Corentin Tallec}, \bibinfo{person}{Pierre~H. Richemond},
  \bibinfo{person}{Elena Buchatskaya}, \bibinfo{person}{Carl Doersch},
  \bibinfo{person}{Bernardo~Avila Pires}, \bibinfo{person}{Zhaohan~Daniel Guo},
  \bibinfo{person}{Mohammad~Gheshlaghi Azar}, \bibinfo{person}{Bilal Piot},
  \bibinfo{person}{Koray Kavukcuoglu}, \bibinfo{person}{R\'{e}mi Munos}, {and}
  \bibinfo{person}{Michal Valko}.} \bibinfo{year}{2020}\natexlab{}.
\newblock \showarticletitle{Bootstrap Your Own Latent a New Approach to
  Self-Supervised Learning}. In \bibinfo{booktitle}{\emph{Proceedings of the
  34th International Conference on Neural Information Processing Systems}}.
  \bibinfo{pages}{21271--21284}.
\newblock


\bibitem[Gutmann and Hyv{\"a}rinen(2010)]%
        {Gutmann:2010:ICAIS}
\bibfield{author}{\bibinfo{person}{Michael Gutmann} {and} \bibinfo{person}{Aapo
  Hyv{\"a}rinen}.} \bibinfo{year}{2010}\natexlab{}.
\newblock \showarticletitle{Noise-contrastive estimation: A new estimation
  principle for unnormalized statistical models}. In
  \bibinfo{booktitle}{\emph{Proceedings of the thirteenth international
  conference on artificial intelligence and statistics}}.
  \bibinfo{pages}{297--304}.
\newblock


\bibitem[He et~al\mbox{.}(2020)]%
        {He:2020:CVPR}
\bibfield{author}{\bibinfo{person}{Kaiming He}, \bibinfo{person}{Haoqi Fan},
  \bibinfo{person}{Yuxin Wu}, \bibinfo{person}{Saining Xie}, {and}
  \bibinfo{person}{Ross Girshick}.} \bibinfo{year}{2020}\natexlab{}.
\newblock \showarticletitle{Momentum contrast for unsupervised visual
  representation learning}. In \bibinfo{booktitle}{\emph{Proceedings of the
  IEEE/CVF conference on Computer Vision and Pattern Recognition}}.
  \bibinfo{pages}{9729--9738}.
\newblock


\bibitem[Hjelm et~al\mbox{.}(2018)]%
        {Hjelm:2018:Arxiv}
\bibfield{author}{\bibinfo{person}{R~Devon Hjelm}, \bibinfo{person}{Alex
  Fedorov}, \bibinfo{person}{Samuel Lavoie-Marchildon}, \bibinfo{person}{Karan
  Grewal}, \bibinfo{person}{Phil Bachman}, \bibinfo{person}{Adam Trischler},
  {and} \bibinfo{person}{Yoshua Bengio}.} \bibinfo{year}{2018}\natexlab{}.
\newblock \showarticletitle{Learning deep representations by mutual information
  estimation and maximization}.
\newblock \bibinfo{journal}{\emph{arXiv preprint arXiv:1808.06670}}
  (\bibinfo{year}{2018}).
\newblock


\bibitem[Huang et~al\mbox{.}(2019)]%
        {Huang:2019:ICML}
\bibfield{author}{\bibinfo{person}{Jiabo Huang}, \bibinfo{person}{Qi Dong},
  \bibinfo{person}{Shaogang Gong}, {and} \bibinfo{person}{Xiatian Zhu}.}
  \bibinfo{year}{2019}\natexlab{}.
\newblock \showarticletitle{Unsupervised deep learning by neighbourhood
  discovery}. In \bibinfo{booktitle}{\emph{International Conference on Machine
  Learning}}. \bibinfo{pages}{2849--2858}.
\newblock


\bibitem[Joshi and Mirzasoleiman(2023)]%
        {joshi2023data}
\bibfield{author}{\bibinfo{person}{Siddharth Joshi} {and}
  \bibinfo{person}{Baharan Mirzasoleiman}.} \bibinfo{year}{2023}\natexlab{}.
\newblock \showarticletitle{Data-Efficient Contrastive Self-supervised
  Learning: Easy Examples Contribute the Most}.
\newblock \bibinfo{journal}{\emph{arXiv preprint arXiv:2302.09195}}
  (\bibinfo{year}{2023}).
\newblock


\bibitem[Koren et~al\mbox{.}(2009)]%
        {Koren:2009:Computer}
\bibfield{author}{\bibinfo{person}{Yehuda Koren}, \bibinfo{person}{Robert
  Bell}, {and} \bibinfo{person}{Chris Volinsky}.}
  \bibinfo{year}{2009}\natexlab{}.
\newblock \showarticletitle{Matrix factorization techniques for recommender
  systems}.
\newblock \bibinfo{journal}{\emph{Computer}} \bibinfo{volume}{42},
  \bibinfo{number}{8} (\bibinfo{year}{2009}), \bibinfo{pages}{30--37}.
\newblock


\bibitem[{Li} et~al\mbox{.}(2021)]%
        {Li:2021:ICLR}
\bibfield{author}{\bibinfo{person}{Junnan {Li}}, \bibinfo{person}{Pan {Zhou}},
  \bibinfo{person}{Caiming {Xiong}}, {and} \bibinfo{person}{Steven C.~H.
  {Hoi}}.} \bibinfo{year}{2021}\natexlab{}.
\newblock \showarticletitle{Prototypical Contrastive Learning of Unsupervised
  Representations}. In \bibinfo{booktitle}{\emph{International Conference on
  Learning Representations}}. \bibinfo{pages}{353--356}.
\newblock


\bibitem[Lin et~al\mbox{.}(2022)]%
        {lin2022improving}
\bibfield{author}{\bibinfo{person}{Zihan Lin}, \bibinfo{person}{Changxin Tian},
  \bibinfo{person}{Yupeng Hou}, {and} \bibinfo{person}{Wayne~Xin Zhao}.}
  \bibinfo{year}{2022}\natexlab{}.
\newblock \showarticletitle{Improving graph collaborative filtering with
  neighborhood-enriched contrastive learning}. In
  \bibinfo{booktitle}{\emph{Proceedings of the ACM Web Conference 2022}}.
  \bibinfo{pages}{2320--2329}.
\newblock


\bibitem[Liu and Wang(2023)]%
        {Bin:2023:ICDE}
\bibfield{author}{\bibinfo{person}{Bin Liu} {and} \bibinfo{person}{Bang Wang}.}
  \bibinfo{year}{2023}\natexlab{}.
\newblock \showarticletitle{Bayesian Negative Sampling for Recommendation}. In
  \bibinfo{booktitle}{\emph{2023 IEEE 39th International Conference on Data
  Engineering (ICDE)}}. IEEE, \bibinfo{pages}{749--761}.
\newblock


\bibitem[Liu et~al\mbox{.}(2021b)]%
        {Liu:2021:TKDE}
\bibfield{author}{\bibinfo{person}{Xiao Liu}, \bibinfo{person}{Fanjin Zhang},
  \bibinfo{person}{Zhenyu Hou}, \bibinfo{person}{Li Mian},
  \bibinfo{person}{Zhaoyu Wang}, \bibinfo{person}{Jing Zhang}, {and}
  \bibinfo{person}{Jie Tang}.} \bibinfo{year}{2021}\natexlab{b}.
\newblock \showarticletitle{Self-supervised learning: Generative or
  contrastive}.
\newblock \bibinfo{journal}{\emph{IEEE Transactions on Knowledge and Data
  Engineering}} (\bibinfo{year}{2021}).
\newblock


\bibitem[Liu et~al\mbox{.}(2021a)]%
        {liu2021contrastive}
\bibfield{author}{\bibinfo{person}{Zhuang Liu}, \bibinfo{person}{Yunpu Ma},
  \bibinfo{person}{Yuanxin Ouyang}, {and} \bibinfo{person}{Zhang Xiong}.}
  \bibinfo{year}{2021}\natexlab{a}.
\newblock \showarticletitle{Contrastive learning for recommender system}.
\newblock \bibinfo{journal}{\emph{arXiv preprint arXiv:2101.01317}}
  (\bibinfo{year}{2021}).
\newblock


\bibitem[Luo et~al\mbox{.}(2023)]%
        {luo2023segclip}
\bibfield{author}{\bibinfo{person}{Huaishao Luo}, \bibinfo{person}{Junwei Bao},
  \bibinfo{person}{Youzheng Wu}, \bibinfo{person}{Xiaodong He}, {and}
  \bibinfo{person}{Tianrui Li}.} \bibinfo{year}{2023}\natexlab{}.
\newblock \showarticletitle{Segclip: Patch aggregation with learnable centers
  for open-vocabulary semantic segmentation}. In
  \bibinfo{booktitle}{\emph{ICML}}. \bibinfo{pages}{23033--23044}.
\newblock


\bibitem[Oord et~al\mbox{.}(2018)]%
        {Oord:2018:arxiv}
\bibfield{author}{\bibinfo{person}{Aaron van~den Oord}, \bibinfo{person}{Yazhe
  Li}, {and} \bibinfo{person}{Orio Vinyals}.} \bibinfo{year}{2018}\natexlab{}.
\newblock \showarticletitle{Representation learning with contrastive predictive
  coding}.
\newblock \bibinfo{journal}{\emph{arXiv preprint arXiv:1807.03748}}
  (\bibinfo{year}{2018}).
\newblock


\bibitem[Qiu et~al\mbox{.}(2022)]%
        {qiu2022contrastive}
\bibfield{author}{\bibinfo{person}{Ruihong Qiu}, \bibinfo{person}{Zi Huang},
  \bibinfo{person}{Hongzhi Yin}, {and} \bibinfo{person}{Zijian Wang}.}
  \bibinfo{year}{2022}\natexlab{}.
\newblock \showarticletitle{Contrastive learning for representation
  degeneration problem in sequential recommendation}. In
  \bibinfo{booktitle}{\emph{Proceedings of the fifteenth ACM international
  conference on web search and data mining}}. \bibinfo{pages}{813--823}.
\newblock


\bibitem[Rendle and Freudenthaler(2014)]%
        {Steffen:2014:WSDM}
\bibfield{author}{\bibinfo{person}{Steffen Rendle} {and}
  \bibinfo{person}{Christoph Freudenthaler}.} \bibinfo{year}{2014}\natexlab{}.
\newblock \showarticletitle{Improving pairwise learning for item recommendation
  from implicit feedback}. In \bibinfo{booktitle}{\emph{Proceedings of the 7th
  ACM international conference on Web Search and Data Mining}}.
  \bibinfo{pages}{273--282}.
\newblock


\bibitem[Rendle et~al\mbox{.}(2009)]%
        {Steffen:2009:UAI}
\bibfield{author}{\bibinfo{person}{Steffen Rendle}, \bibinfo{person}{Christoph
  Freudenthaler}, \bibinfo{person}{Zeno Gantner}, {and} \bibinfo{person}{Lars
  Schmidt{-}Thieme}.} \bibinfo{year}{2009}\natexlab{}.
\newblock \showarticletitle{{BPR:} Bayesian Personalized Ranking from Implicit
  Feedback}. In \bibinfo{booktitle}{\emph{{UAI} 2009, Proceedings of the
  Twenty-Fifth Conference on Uncertainty in Artificial Intelligence, Montreal,
  QC, Canada, June 18-21, 2009}}. \bibinfo{pages}{452--461}.
\newblock


\bibitem[Robinson et~al\mbox{.}(2021)]%
        {Robinson:2021:ICLR}
\bibfield{author}{\bibinfo{person}{Joshua Robinson}, \bibinfo{person}{Chuang
  Ching-Yao}, \bibinfo{person}{Suvrit Sra}, {and} \bibinfo{person}{Stefanie
  Jegelka}.} \bibinfo{year}{2021}\natexlab{}.
\newblock \showarticletitle{Contrastive Learning with Hard Negative Samples}.
  In \bibinfo{booktitle}{\emph{International Conference on Learning
  Representations}}.
\newblock


\bibitem[Sun et~al\mbox{.}(2019)]%
        {sun2019bert4rec}
\bibfield{author}{\bibinfo{person}{Fei Sun}, \bibinfo{person}{Jun Liu},
  \bibinfo{person}{Jian Wu}, \bibinfo{person}{Changhua Pei},
  \bibinfo{person}{Xiao Lin}, \bibinfo{person}{Wenwu Ou}, {and}
  \bibinfo{person}{Peng Jiang}.} \bibinfo{year}{2019}\natexlab{}.
\newblock \showarticletitle{BERT4Rec: Sequential recommendation with
  bidirectional encoder representations from transformer}. In
  \bibinfo{booktitle}{\emph{Proceedings of the 28th ACM international
  conference on information and knowledge management}}.
  \bibinfo{pages}{1441--1450}.
\newblock


\bibitem[Wang et~al\mbox{.}(2023)]%
        {wang2023curriculum}
\bibfield{author}{\bibinfo{person}{Hui Wang}, \bibinfo{person}{Kun Zhou},
  \bibinfo{person}{Xin Zhao}, \bibinfo{person}{Jingyuan Wang}, {and}
  \bibinfo{person}{Ji-Rong Wen}.} \bibinfo{year}{2023}\natexlab{}.
\newblock \showarticletitle{Curriculum pre-training heterogeneous subgraph
  transformer for top-n recommendation}.
\newblock \bibinfo{journal}{\emph{ACM Transactions on Information Systems}}
  \bibinfo{volume}{41}, \bibinfo{number}{1} (\bibinfo{year}{2023}),
  \bibinfo{pages}{1--28}.
\newblock


\bibitem[Wang and Isola(2020)]%
        {Wang:2020:ICML}
\bibfield{author}{\bibinfo{person}{Tongzhou Wang} {and}
  \bibinfo{person}{Phillip Isola}.} \bibinfo{year}{2020}\natexlab{}.
\newblock \showarticletitle{Understanding contrastive representation learning
  through alignment and uniformity on the hypersphere}. In
  \bibinfo{booktitle}{\emph{International Conference on Machine Learning}}.
  \bibinfo{pages}{9929--9939}.
\newblock


\bibitem[Wang et~al\mbox{.}(2021)]%
        {Wang:2021:WSDM}
\bibfield{author}{\bibinfo{person}{Wenjie Wang}, \bibinfo{person}{Fuli Feng},
  \bibinfo{person}{Xiangnan He}, \bibinfo{person}{Liqiang Nie}, {and}
  \bibinfo{person}{Tat-Seng Chua}.} \bibinfo{year}{2021}\natexlab{}.
\newblock \showarticletitle{Denoising implicit feedback for recommendation}. In
  \bibinfo{booktitle}{\emph{Proceedings of the 14th ACM international
  conference on web search and data mining}}. \bibinfo{pages}{373--381}.
\newblock


\bibitem[Wang et~al\mbox{.}(2019)]%
        {Wang:2019:SIGIR}
\bibfield{author}{\bibinfo{person}{Xiang Wang}, \bibinfo{person}{Xiangnan He},
  \bibinfo{person}{Meng Wang}, \bibinfo{person}{Fuli Feng}, {and}
  \bibinfo{person}{Tat~Seng Chua}.} \bibinfo{year}{2019}\natexlab{}.
\newblock \showarticletitle{Neural Graph Collaborative Filtering}. In
  \bibinfo{booktitle}{\emph{Proceedings of the International ACM SIGIR
  Conference on Research and Development in Information Retrieval}}.
  \bibinfo{pages}{2344--2353}.
\newblock


\bibitem[Wenqi et~al\mbox{.}(2021)]%
        {Wenqi:2021:KDD}
\bibfield{author}{\bibinfo{person}{Fan Wenqi}, \bibinfo{person}{Liu Xiaorui},
  \bibinfo{person}{Jin Wei}, \bibinfo{person}{Zhao Xiangyu},
  \bibinfo{person}{Tang Jiliang}, {and} \bibinfo{person}{Li Qing}.}
  \bibinfo{year}{2021}\natexlab{}.
\newblock \showarticletitle{Graph Trend Networks for Recommendations}. In
  \bibinfo{booktitle}{\emph{Proceedings of the ACM SIGKDD Conference on
  Knowledge Discovery and Data Mining}}. \bibinfo{pages}{12}.
\newblock


\bibitem[Wu et~al\mbox{.}(2021)]%
        {10.1145/3404835.3462862}
\bibfield{author}{\bibinfo{person}{Jiancan Wu}, \bibinfo{person}{Xiang Wang},
  \bibinfo{person}{Fuli Feng}, \bibinfo{person}{Xiangnan He},
  \bibinfo{person}{Liang Chen}, \bibinfo{person}{Jianxun Lian}, {and}
  \bibinfo{person}{Xing Xie}.} \bibinfo{year}{2021}\natexlab{}.
\newblock \showarticletitle{Self-Supervised Graph Learning for Recommendation}.
  In \bibinfo{booktitle}{\emph{Proceedings of the 44th International ACM SIGIR
  Conference on Research and Development in Information Retrieval}}.
  \bibinfo{pages}{726–735}.
\newblock


\bibitem[Wu et~al\mbox{.}(2022)]%
        {Jiancan:2022:arxiv}
\bibfield{author}{\bibinfo{person}{Jiancan Wu}, \bibinfo{person}{Xiang Wang},
  \bibinfo{person}{Xingyu Gao}, \bibinfo{person}{Jiawei Chen},
  \bibinfo{person}{Hongcheng Fu}, {and} \bibinfo{person}{Tianyu~Qiu andXiangnan
  He}.} \bibinfo{year}{2022}\natexlab{}.
\newblock \showarticletitle{On the Effectiveness of Sampled Softmax Loss for
  Item Recommendation}.
\newblock \bibinfo{journal}{\emph{arXiv preprint arXiv:2201.02327}}
  (\bibinfo{year}{2022}).
\newblock


\bibitem[Wu et~al\mbox{.}(2023)]%
        {wu:2023:TKDE}
\bibfield{author}{\bibinfo{person}{Lirong Wu}, \bibinfo{person}{Haitao Lin},
  \bibinfo{person}{Cheng Tan}, \bibinfo{person}{Zhangyang Gao}, {and}
  \bibinfo{person}{Stan~Z. Li}.} \bibinfo{year}{2023}\natexlab{}.
\newblock \showarticletitle{Self-Supervised Learning on Graphs: Contrastive,
  Generative, or Predictive}.
\newblock \bibinfo{journal}{\emph{IEEE Transactions on Knowledge and Data
  Engineering}} \bibinfo{volume}{35}, \bibinfo{number}{4}
  (\bibinfo{year}{2023}), \bibinfo{pages}{4216--4235}.
\newblock


\bibitem[Wu et~al\mbox{.}(2018)]%
        {Wu:2018:CVPR}
\bibfield{author}{\bibinfo{person}{Zhirong Wu}, \bibinfo{person}{Yuanjun
  Xiong}, \bibinfo{person}{Stella~X Yu}, {and} \bibinfo{person}{Dahua Lin}.}
  \bibinfo{year}{2018}\natexlab{}.
\newblock \showarticletitle{Unsupervised feature learning via non-parametric
  instance discrimination}. In \bibinfo{booktitle}{\emph{Proceedings of the
  IEEE/CVF conference on Computer Vision and Pattern Recognition}}.
  \bibinfo{pages}{3733--3742}.
\newblock


\bibitem[Xiangnan et~al\mbox{.}(2020)]%
        {Xiangnan:2020:SIGIR}
\bibfield{author}{\bibinfo{person}{He Xiangnan}, \bibinfo{person}{Deng Kuan},
  \bibinfo{person}{Wang Xiang}, \bibinfo{person}{Li Yan},
  \bibinfo{person}{Zhang Yongdong}, {and} \bibinfo{person}{Wang Meng}.}
  \bibinfo{year}{2020}\natexlab{}.
\newblock \showarticletitle{LightGCN: Simplifying and Powering Graph
  Convolution Network for Recommendation.}. In
  \bibinfo{booktitle}{\emph{Proceedings of the International ACM SIGIR
  Conference on Research and Development in Information Retrieval}}.
  \bibinfo{pages}{10}.
\newblock


\bibitem[Yang et~al\mbox{.}(2022)]%
        {10.1145/3477495.3532009}
\bibfield{author}{\bibinfo{person}{Yuhao Yang}, \bibinfo{person}{Chao Huang},
  \bibinfo{person}{Lianghao Xia}, {and} \bibinfo{person}{Chenliang Li}.}
  \bibinfo{year}{2022}\natexlab{}.
\newblock \showarticletitle{Knowledge Graph Contrastive Learning for
  Recommendation}. In \bibinfo{booktitle}{\emph{Proceedings of the 45th
  International ACM SIGIR Conference on Research and Development in Information
  Retrieval}}. \bibinfo{pages}{1434–1443}.
\newblock


\bibitem[Yang et~al\mbox{.}(2020)]%
        {Yang:2020:KDD}
\bibfield{author}{\bibinfo{person}{Zhen Yang}, \bibinfo{person}{Ming Ding},
  \bibinfo{person}{Chang Zhou}, \bibinfo{person}{Hongxia Yang},
  \bibinfo{person}{Jingren Zhou}, {and} \bibinfo{person}{Jie Tang}.}
  \bibinfo{year}{2020}\natexlab{}.
\newblock \showarticletitle{Understanding negative sampling in graph
  representation learning}. In \bibinfo{booktitle}{\emph{Proceedings of the
  26th ACM SIGKDD International Conference on Knowledge Discovery \& Data
  Mining}}. \bibinfo{pages}{1666--1676}.
\newblock


\bibitem[Yu et~al\mbox{.}(2023a)]%
        {yu2023xsimgcl}
\bibfield{author}{\bibinfo{person}{Junliang Yu}, \bibinfo{person}{Xin Xia},
  \bibinfo{person}{Tong Chen}, \bibinfo{person}{Lizhen Cui},
  \bibinfo{person}{Nguyen Quoc~Viet Hung}, {and} \bibinfo{person}{Hongzhi
  Yin}.} \bibinfo{year}{2023}\natexlab{a}.
\newblock \showarticletitle{XSimGCL: Towards extremely simple graph contrastive
  learning for recommendation}.
\newblock \bibinfo{journal}{\emph{IEEE Transactions on Knowledge and Data
  Engineering}} (\bibinfo{year}{2023}).
\newblock


\bibitem[Yu et~al\mbox{.}(2023b)]%
        {SSR:2023:TKDE}
\bibfield{author}{\bibinfo{person}{Junliang Yu}, \bibinfo{person}{Hongzhi Yin},
  \bibinfo{person}{Xin Xia}, \bibinfo{person}{Tong Chen},
  \bibinfo{person}{Jundong Li}, {and} \bibinfo{person}{Zi Huang}.}
  \bibinfo{year}{2023}\natexlab{b}.
\newblock \showarticletitle{Self-Supervised Learning for Recommender Systems: A
  Survey}.
\newblock \bibinfo{journal}{\emph{IEEE Transactions on Knowledge and Data
  Engineering}} (\bibinfo{year}{2023}), \bibinfo{pages}{1--20}.
\newblock
\showISSN{1558-2191}


\bibitem[Zhang et~al\mbox{.}(2021)]%
        {zhang2021understanding}
\bibfield{author}{\bibinfo{person}{Chiyuan Zhang}, \bibinfo{person}{Samy
  Bengio}, \bibinfo{person}{Moritz Hardt}, \bibinfo{person}{Benjamin Recht},
  {and} \bibinfo{person}{Oriol Vinyals}.} \bibinfo{year}{2021}\natexlab{}.
\newblock \showarticletitle{Understanding deep learning (still) requires
  rethinking generalization}.
\newblock \bibinfo{journal}{\emph{Commun. ACM}} \bibinfo{volume}{64},
  \bibinfo{number}{3} (\bibinfo{year}{2021}), \bibinfo{pages}{107--115}.
\newblock


\bibitem[Zhang et~al\mbox{.}(2013)]%
        {Zhang:2013:SIGIR}
\bibfield{author}{\bibinfo{person}{Weinan Zhang}, \bibinfo{person}{Tianqi
  Chen}, \bibinfo{person}{Jun Wang}, {and} \bibinfo{person}{Yong Yu}.}
  \bibinfo{year}{2013}\natexlab{}.
\newblock \showarticletitle{Optimizing Top-n Collaborative Filtering via
  Dynamic Negative Item Sampling}. In \bibinfo{booktitle}{\emph{Proceedings of
  the 36th International ACM SIGIR Conference on Research and Development in
  Information Retrieval}}. \bibinfo{pages}{785–788}.
\newblock


\bibitem[Zhang and Wang(2023)]%
        {zhang:sigir}
\bibfield{author}{\bibinfo{person}{Zizhuo Zhang} {and} \bibinfo{person}{Bang
  Wang}.} \bibinfo{year}{2023}\natexlab{}.
\newblock \showarticletitle{Prompt Learning for News Recommendation}. In
  \bibinfo{booktitle}{\emph{Proceedings of the 46th International ACM SIGIR
  Conference on Research and Development in Information Retrieval}}.
  \bibinfo{pages}{227–237}.
\newblock


\bibitem[Zhou et~al\mbox{.}(2023)]%
        {10.1145/3543507.3583251}
\bibfield{author}{\bibinfo{person}{Xin Zhou}, \bibinfo{person}{Hongyu Zhou},
  \bibinfo{person}{Yong Liu}, \bibinfo{person}{Zhiwei Zeng},
  \bibinfo{person}{Chunyan Miao}, \bibinfo{person}{Pengwei Wang},
  \bibinfo{person}{Yuan You}, {and} \bibinfo{person}{Feijun Jiang}.}
  \bibinfo{year}{2023}\natexlab{}.
\newblock \showarticletitle{Bootstrap Latent Representations for Multi-Modal
  Recommendation}. In \bibinfo{booktitle}{\emph{Proceedings of the ACM Web
  Conference 2023}}. \bibinfo{pages}{845–854}.
\newblock
\showISBNx{9781450394161}


\end{thebibliography}

\end{document}